\documentclass[%
 aip,
 amsmath,amssymb,
 reprint,%
]{revtex4-2}
\usepackage{array}
\usepackage{textcomp}
\usepackage{chemformula}
\usepackage{graphicx}
\usepackage{dcolumn}
\usepackage{bm}
\usepackage{changes}
\usepackage{xcolor}
\usepackage{makecell}
\usepackage{blindtext}

\usepackage[T1]{fontenc}
\usepackage{mathptmx}

\begin{document}

\preprint{AIP/123-QED}

\title[Self-sealing complex oxide resonators]{Self-sealing complex oxide resonators}
\author{Martin Lee}
\affiliation{Kavli Institute of Nanoscience, Delft University of Technology, Lorentzweg 1, 2628 CJ Delft, The Netherlands.}

\author{Martin Robin}
\affiliation{Department of Precision and Microsystems Engineering, Delft University of Technology, Mekelweg 2, 2628 CD Delft, The Netherlands.}

\author{Ruben Guis}
\affiliation{Department of Precision and Microsystems Engineering, Delft University of Technology, Mekelweg 2, 2628 CD Delft, The Netherlands.}

\author{Ulderico Filippozzi}
\affiliation{Kavli Institute of Nanoscience, Delft University of Technology, Lorentzweg 1, 2628 CJ Delft, The Netherlands.}

\author{Dong Hoon Shin}
\affiliation{Kavli Institute of Nanoscience, Delft University of Technology, Lorentzweg 1, 2628 CJ Delft, The Netherlands.}

\author{Thierry C. van Thiel}
\affiliation{Kavli Institute of Nanoscience, Delft University of Technology, Lorentzweg 1, 2628 CJ Delft, The Netherlands.}

\author{Stijn Paardekooper}
\affiliation{Department of Precision and Microsystems Engineering, Delft University of Technology, Mekelweg 2, 2628 CD Delft, The Netherlands.}

\author{Johannes R. Renshof}
\affiliation{Kavli Institute of Nanoscience, Delft University of Technology, Lorentzweg 1, 2628 CJ Delft, The Netherlands.}

\author{Herre S. J. van der Zant}
\affiliation{Kavli Institute of Nanoscience, Delft University of Technology, Lorentzweg 1, 2628 CJ Delft, The Netherlands.}

\author{Andrea D. Caviglia}
\affiliation{Kavli Institute of Nanoscience, Delft University of Technology, Lorentzweg 1, 2628 CJ Delft, The Netherlands.}

\author{Gerard J. Verbiest}
\affiliation{Department of Precision and Microsystems Engineering, Delft University of Technology, Mekelweg 2, 2628 CD Delft, The Netherlands.}

\author{Peter G. Steeneken}
\affiliation{Kavli Institute of Nanoscience, Delft University of Technology, Lorentzweg 1, 2628 CJ Delft, The Netherlands.}
\affiliation{Department of Precision and Microsystems Engineering, Delft University of Technology, Mekelweg 2, 2628 CD Delft, The Netherlands.}

\date{\today}

\begin{abstract}

Although 2D materials hold great potential for next-generation pressure sensors, recent studies revealed that gases permeate along the membrane-surface interface that is only weakly bound by van der Waals interactions, necessitating additional sealing procedures.
In this work, we demonstrate the use of free-standing complex oxides as self-sealing membranes that allow the reference cavity of pressure sensors to be sealed by a simple anneal. 
To test the hermeticity, we study the gas permeation time constants in nano-mechanical resonators made from \ch{SrRuO3} and \ch{SrTiO3} membranes suspended over \ch{SiO2}/Si cavities which show an improvement up to 4 orders of magnitude in the permeation time constant after annealing the devices for 15 minutes. Similar devices fabricated on \ch{Si3N4/Si} do not show such improvements, suggesting that the adhesion increase over \ch{SiO2} is mediated by oxygen bonds that are formed at the \ch{SiO2}/complex oxide interface during the self-sealing anneal. 
We confirm the enhancement of adhesion by picosecond ultrasonics measurements which show an increase in the interfacial stiffness by 70\% after annealing. Since it is straigthforward to apply, the presented self-sealing method is thus a promising route toward realizing ultrathin hermetic pressure sensors.

\end{abstract}

\maketitle

\section{Introduction}

Van der Waals (vdW) materials attracted significant attention in the microelectromechanical systems (MEMS) community due to their low dimensionality, flexibility and strength \cite{lemme2020nanoelectromechanical}. In particular, graphene is considered as the material for the next generation pressure sensors \cite{dolleman2016graphene,smith2013electromechanical,smith2016piezoresistive,zhu2013graphene,vsivskins2020sensitive,davidovikj2017static} thanks to its intrinsic impermeability to gases \cite{berry2013impermeability,bunch2008impermeable,sun2020limits}. Pressure sensors operate by measuring the deflection of a membrane due to the pressure difference between a reference cavity and the environment. For a reliable pressure sensor, hermeticity of the cavity underneath the membrane is essential. However, gas permeation along the interface between the vdW membrane and the substrate causes pressure variations in the reference cavity, which renders the pressure readings from graphene-based pressure sensors unreliable  \cite{lee2019sealing,manzanares2020improved}. 

Recently reported sealing protocols have enabled improvements in the hermeticity of vdW material membranes of up to a factor 10000 \cite{lee2019sealing,manzanares2020improved}, but
scaling them to high volume production is difficult, since depositing and patterning of sealing layers on top of ultrathin vdW material membranes is often detrimental to device performance, in particular if high temperatures are needed. Moreover, the pressure at which the sealing layer is deposited is often fixed by the process, such that the reference pressure in the cavity cannot be freely controlled \cite{wunnicke2011small, partridge2005mems, partridge2001new, seetharaman2010robust}.


As an alternative to graphene, we introduce in this letter, free-standing single crystal complex oxide perovskites as a membrane for pressure sensing MEMS applications. Thanks to their inter-unit cell chemical bonds and the ability to form new ones at high temperatures \cite{yong2010thermal}, they promise a stronger adhesion to the substrate than 2D materials which are mediated by a vdW gap. Moreover, due to their epitaxial crystalline growth, extremely flat layer surfaces can be grown using pulsed laser deposition, enhancing the interface contact and reducing the formation probability of gas leakage pathways. Recent developments in releasing epitaxially grown single crystal complex oxides allow them to be thinned down to the unit cell limit, similar to the vdW materials \cite{lu2016synthesis,ji2019freestanding,kum2020heterogeneous}. Complex oxides in their ultra-thin free-standing form are mechanically robust \cite{harbola2021strain} while withstanding strains up to 8\% \cite{hong2020extreme,xu2020strain}, are flexible enough to allow large curvatures \cite{peng2020phase} and have already been demonstrated as viable nanomechanical resonators \cite{davidovikj2020ultrathin,manca2019large}. Furthermore, wafer-scale production of single crystalline complex oxides are being developed \cite{blank2013pulsed} which makes them even more appealing for large-scale CMOS compatible fabrication.

Here, we use free-standing \ch{SrRuO3} (SRO) and \ch{SrTiO3} (STO) suspended over \ch{SiO2/Si} cavities to make pressure sensors and demonstrate a simple, CMOS compatible sealing technique which does not require additional fabrication steps. The sealing consists of annealing the devices above 300 $^\circ$C in ambient conditions for 15 minutes. We measure the time dependence of the resonance frequency to extract the gas permeation time constant. By comparing the permeation time constant of the pressure sensor devices before and after performing the self-sealing annealing process, we show that the permeation time constant increases from 14 s to $>$10000 s, indicative of a large increase in hermeticity. Comparable devices fabricated on \ch{Si3N4/Si} cavities do not show such enhancement of the hermeticity which suggests that the improved hermeticity is mediated by the properties of the \ch{SiO2} that promote adhesion to the complex oxides, thus eliminating gas leakage rates.
Furthermore, we probe the STO-\ch{SiO2} interface in both annealed and non-annealed samples using a picosecond ultrasonics technique. The measurements show a clear reduction in the ultrasonic reflection coefficient at the interface between STO-\ch{SiO2} after the annealing procedure, indicative of an increased adhesion.
Our work investigates the use of ultrathin complex oxide membranes for pressure sensors, and demonstrates self-sealing of the interface after transfer, thus providing an alternative to graphene and MEMS sensor technologies.

\begin{figure}[htb!]
\includegraphics[width=1\columnwidth]{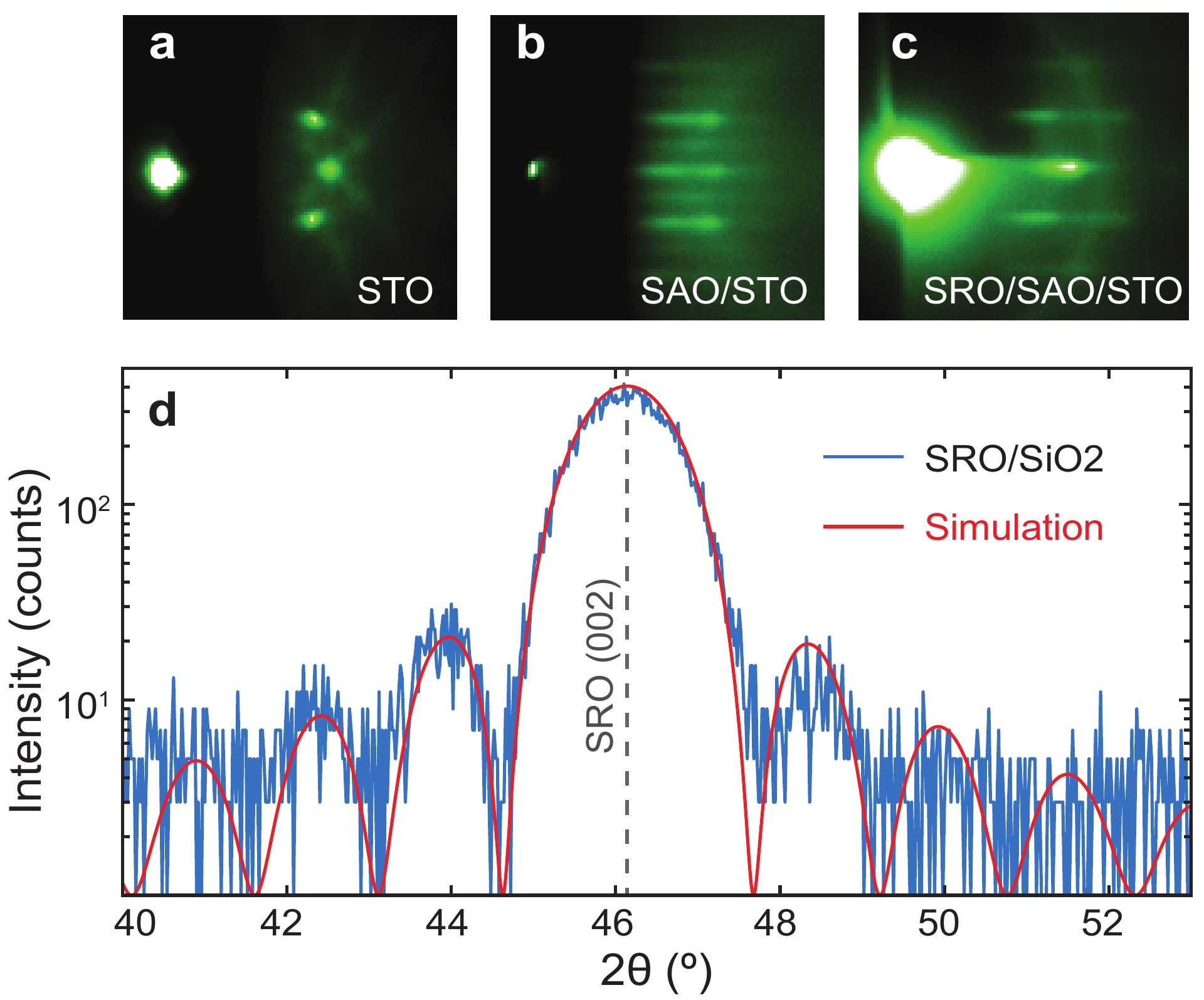}
\caption{Reflection high energy electron diffraction (RHEED) images of \textbf{a} \ch{SrTiO3} (STO) substrate, \textbf{b} \ch{Sr3Al2O6} (SAO) grown on STO substrate and \textbf{c} \ch{SrRuO3} grown on SAO/STO stack. \textbf{d} X-Ray diffraction (blue) of exfoliated SRO stamped on \ch{SiO2/Si} and the simulation(red). The c-axis lattice parameter extracted from the simulation is 3.931 {\AA} and the thickness is 16 unit cells.}
\label{fig:growth}
\end{figure}

\section{Results}
Crystalline free-standing complex oxides are synthesized using pulsed-laser deposition by growing a (water-soluble) buffer layer of \ch{Sr3Al2O6} (SAO) on \ch{SrTiO3} (001) substrates, followed by an overlayer of choice (STO or SRO).  The growth is monitored by in-situ reflection high energy electron diffraction (RHEED), confirming 2D growth (Fig. \ref{fig:growth}a-c). After growth, the samples are attached to polydimethyl siloxane (PDMS) films for support during the etching process of the buffer layer which is performed by submerging the PDMS covered sample in deionized water for 24 hours. After the SAO is etched away, the film of choice can be transferred onto a dummy \ch{SiO2/Si} substrate using a deterministic transfer method \cite{castellanos2014deterministic} for characterization. See supplementary information (S.I-III).

We perform X-ray diffraction (XRD) measurements on the films after transferring parts of them to a dummy \ch{SiO2/Si} to verify the film thicknesses and the crystal coherence. As shown in Fig. \ref{fig:growth}d, the crystallographic (002) peak of SRO can be identified with finite size oscillations on both sides of the main peak, showing long-range crystal coherence of the film after exfoliation and the transfer process. A model fit plotted in red on top of the XRD data, is used to extract the c-axis lattice parameter as well as the number of pseudo-cubic unit cells. In the case of SRO (Fig. \ref{fig:growth}d), the model yields a thickness of 16 unit cells (u.c.) with a c-axis lattice parameter of 3.931 {\AA}, in good agreement with the value reported in the literature \cite{koster2012structure}. After confirming the crystallinity and the thickness of the films, we transfer individual flakes of SRO (6.3 nm) and STO (82 nm) on top of a pre-patterned \ch{SiO2/Si} substrate with circular cavities with diameters from 3 \textmu m to 10 \textmu m, using the vdW pick up technique \cite{pizzocchero2016hot,kim2016van}. Cavities are defined in thermally grown \ch{SiO2} of 285 nm in thickness using e-beam lithography followed by reactive ion etching down to the Si layer \cite{lee2019sealing}.

\begin{figure*}[htb!]
\includegraphics[width=0.8\textwidth]{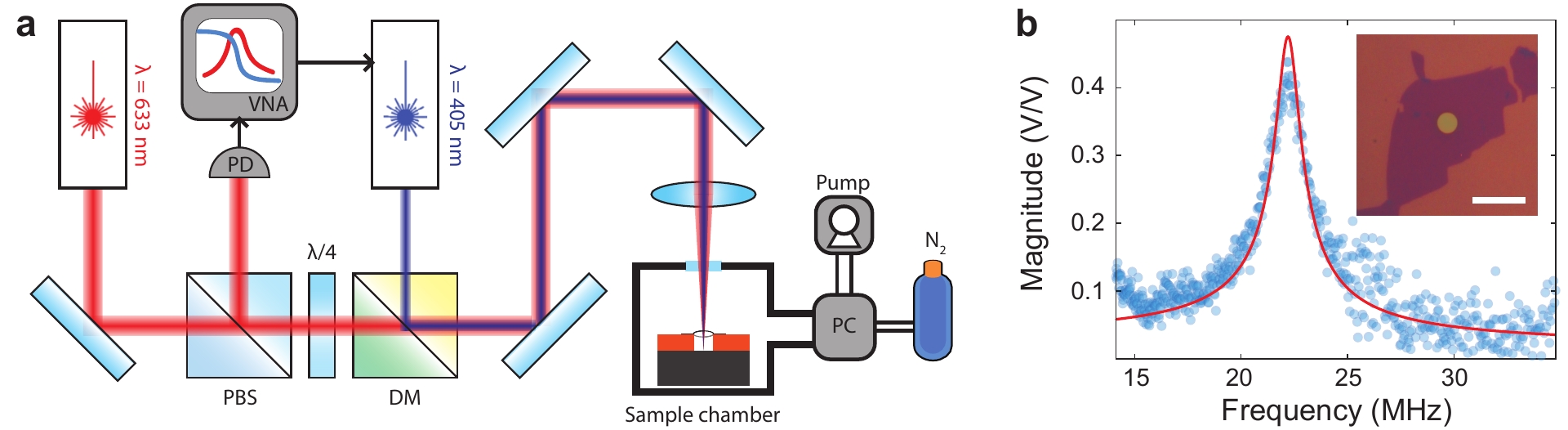}
\caption{\textbf{a} Schematic illustration of the measurement setup. Vector network analyzer (VNA) sends an amplitude modulated signal to the blue laser diode which optothermally actuates the membrane while the red He-Ne laser reads out its motion. The reflected red laser light is detected at the photodetector (PD) and the signal is collected by the VNA. The pressure inside the sample chamber is controlled by the pressure controller (PC) which is connected to a scroll pump and a pressurized \ch{N2} gas bottle. PBS: Polarized Beam Splitter, DM: Dichroic Mirror. \textbf{b} An example of a resonance peak of a SRO (16 u.c.) device with a harmonic oscillator fit in red. Inset: optical image of the device. A SRO flake is stamped on top of a circular cavity in \ch{SiO2/Si}. Scale bar is 10 \textmu m. }
\label{fig:setup}
\end{figure*}

Once the fabrication of suspended complex oxide membranes are completed, we measure the pressure dependence of the resonance frequencies using a laser interferometry technique as illustrated in Fig. \ref{fig:setup}a. An intensity modulated blue ($\lambda$ = 405 nm) laser excites the motion of the membrane which is situated in a pressure controlled environment. A continuous red laser ($\lambda$ = 633 nm) monitors the movement of the membrane. The reflected signal is collected by a photo detector (PD) and the signal is sent to a vector network analyser (VNA). Figure \ref{fig:setup}b shows an example resonance peak of a SRO flake suspended over a circular \ch{SiO2/Si} cavity (see inset). A harmonic oscillator function is fitted to the data (red) which is used to extract the resonance frequency as a function of sample chamber (SC) pressure.

\begin{figure*}[htb!]
\includegraphics[width=\textwidth]{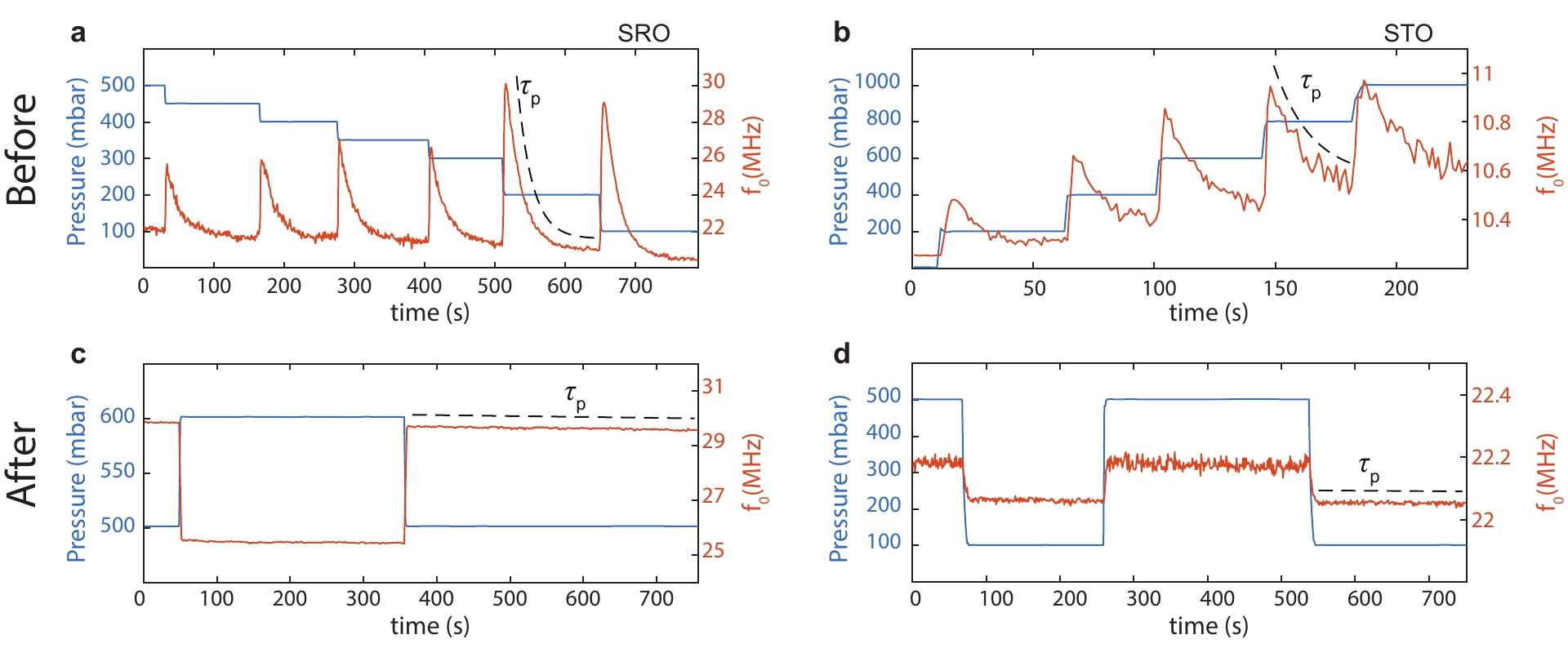}
\caption{Pressure response of the resonance frequency before (\textbf{a}\&\textbf{b}) and after (\textbf{c}\&\textbf{d}) annealing. Left column shows the behavior of a 16 unit cell (6.3 nm) SRO device and the right column shows the behavior of 82 nm STO device. The external pressure controlled by the pressure controller is plotted in blue on the left y-axes and the resonance frequency is plotted in orange on the right y-axes.}
\label{fig:anneal}
\end{figure*}

The time dependent resonance frequency of SRO and STO devices directly after transfer over the cavities are shown in Fig. \ref{fig:anneal}a-b. The SC pressures are adjusted in a step-wise fashion while the frequency is swept to capture the resonance peak. After fitting the data to a harmonic oscillator function, the resonance frequency is extracted and plotted (orange, right y-axis). Before the sealing procedure, both SRO and STO membranes show, after the pressure changes, sudden increases in the resonance frequencies followed by exponential decays of time constant $\tau_{p}$. This behavior suggests that the membranes are tensioned due to the change in the pressure difference in and outside of the cavity, which then quickly equilibrates due to the permeation of gas molecules. In both SRO and STO, the average permeation time constants $\tau_{p}$ are approximately 21 s and 14 s respectively (see S.IV for details on the analysis). 

To reduce the gas leakage, a self-sealing procedure was performed, in which the samples are annealed in air at atmospheric pressure at elevated temperature (for 1 hour at 300 $^\circ$C for SRO and for 15 minutes at 400 $^\circ$C for STO). After this procedure, the measurements from Fig. \ref{fig:anneal}a,b are repeated for both SRO and STO samples and shown in Fig. \ref{fig:anneal}c and \ref{fig:anneal}d respectively. The sudden spike in the resonance frequency followed by a fast decay is not observed in Fig. \ref{fig:anneal}c but instead, a slow reduction in the resonance frequency can be seen. By fitting an exponential decay to the slow reduction in the resonance frequency, we find a $\tau_{p}$ of 1.1 $\times$ 10$^4$ s. Similar behaviour is observed in the STO device after annealing for 15 minutes at 400 $^\circ$C. Before annealing, the mean permeation time constants of STO is $\tau_{p}$ = 13.9 s, which increases to 1.2 $\times$ 10$^5$ s after a self-sealing procedure. As shown in Fig. \ref{fig:anneal}d, no observable decay in the resonance frequencies is present but there are small drifts. It is worth mentioning that due to the minute variations in the resonance frequency, fitting the data from Fig. \ref{fig:anneal}c\&d to an exponential decay is difficult and results in large fit errors. However, we estimate a lower bound of 10$^3$ s.

\begin{figure}[htb!]
\includegraphics[width=\columnwidth]{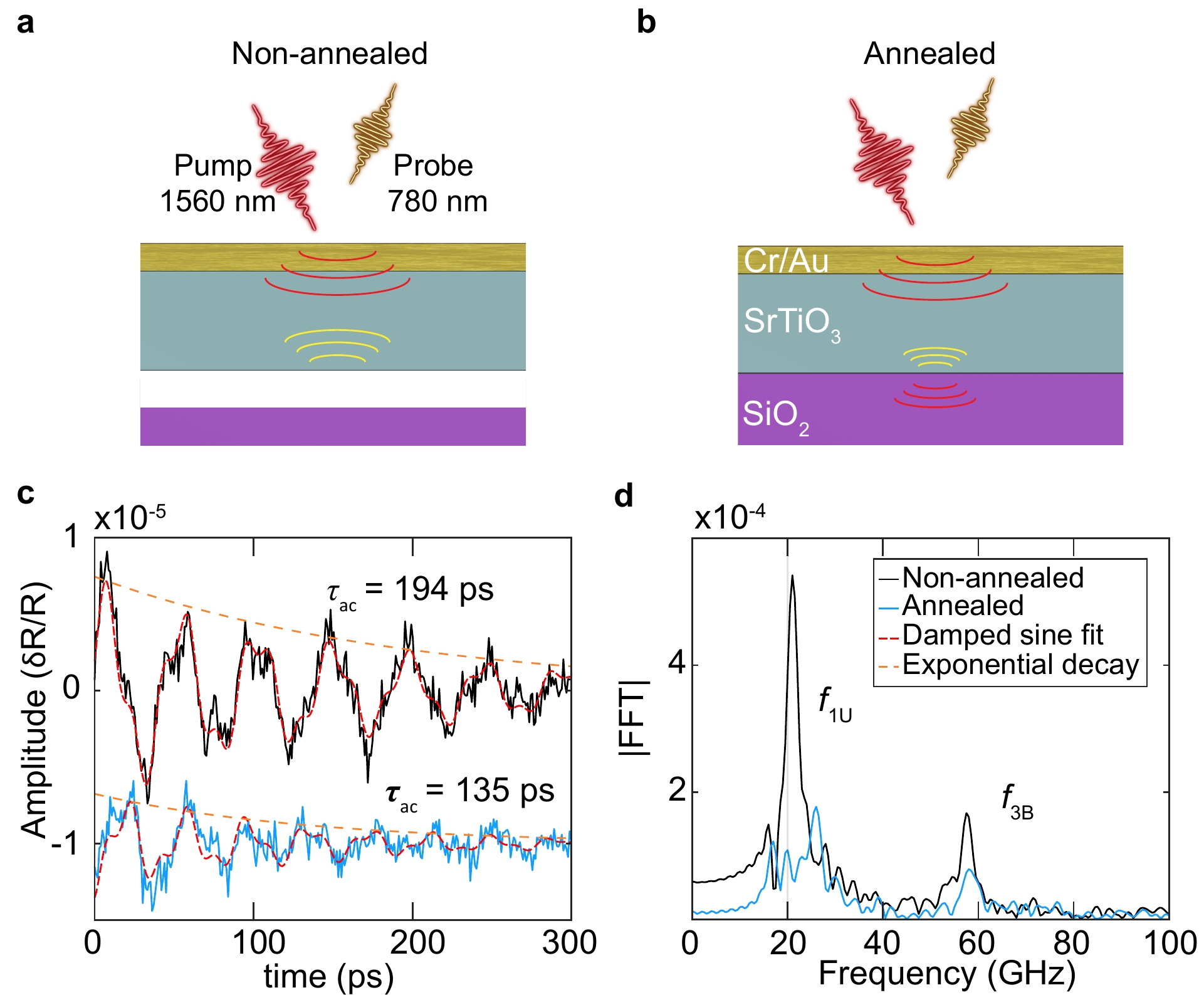}
\caption{Cross sectional illustration of the pump-probe measurement in \textbf{a} non-annealed STO sample and \textbf{b} annealed STO sample. The red acoustic waves depict the propagating wave from the pump pulse and yellow from the reflection at the interface. A 33 nm thick metal layer is deposited on top for the ultra-fast pump-probe measurements. \textbf{c} Examples of picosecond ultrasonics measurements on non-annealed (black) and annealed (light blue, offset in y for easier visualization) flakes of STO. Dashed red lines are fits to the damped sine function and the dashed orange lines depict the exponential decay envelopes. The y-axis shows the relative change in the optical reflection coefficient ($\delta$ R/R) of the probe pulse as a function of the time difference (x-axis) between pump and probe pulse. \textbf{d} Fourier transform of the waves in \textbf{c}.}
\label{fig:pump_probe}
\end{figure}



Next, the adhesion at the interface of complex oxides and the substrate is further tested using a picosecond ultrasonics method. An ultra-fast optical pump-probe setup is used to generate and detect GHz acoustic waves in solids. This allows characterization of the adhesion between the thin layer and the substrate, since these waves are sensitive to the boundary conditions at the interface between two different materials \cite{matsuda2015fundamentals, chang2015tuning, devos2018thin}. For example, this method is used to probe the adhesion properties of metal layers evaporated on glass surfaces \cite{chang2015tuning} or to characterize the adhesion of vdW materials \cite{greener2019high}.

We prepare two sets of flakes on \ch{SiO2/Si} transferred from the same batch of STO (thickness 82 nm). One set is untreated (Fig. \ref{fig:pump_probe}a) while the other is treated with the self-sealing procedure for 1 hour at 400 $^\circ$C (Fig. \ref{fig:pump_probe}b). Both substrates containing STO flakes are then coated with 33 nm of Au/Cr for optical pumping and probing (see S.XI for details). We use an asynchronous optical sampling (ASOPS) technique with a 1560 nm pump laser pulse and a 780 nm probe pulse laser to optothermomechanically generate and detect acoustic echos into individual STO flakes on timescales ranging from 1 ps to 10 ns. The duration of the pulses of both lasers is around 100 fs. Measurements are performed on 4 non-annealed and 5 annealed flakes. An example of the  acoustic measurements on a non-annealed (black) and an annealed (light blue) STO flakes is presented in Fig. \ref{fig:pump_probe}c-d. The overall results are reported in Table \ref{table:results}.

The acoustic wave echos inside the Au/Cr/STO assembly can be seen for the non-annealed and annealed cases in Fig. \ref{fig:pump_probe}c. The decay of the amplitudes of the measurements in the different flakes (Fig. \ref{fig:pump_probe}c for example) are fitted with a damped sine-wave to obtain the time constant $\tau_{ac}$ of the envelope, which characterizes the decay rate of the waves due to reflections at the interface with the substrate, wherein acoustic energy is transmitted. This results in an average value of $\tau_{ac}$ = 220 ps for the non-annealed flakes and $\tau_{ac}$ = 114 ps for the annealed flakes (see Table \ref{table:results}). From these values of $\tau_{ac}$, we calculate the associated acoustic reflection coefficient \cite{greener2019high} $|R_{ac} |$, at the STO/\ch{SiO2} interface. An average value of $|R_{ac} |$ = 0.81 is found for the non-annealed flakes and $|R_{ac} |$ = 0.70 for the annealed flakes. These reflection coefficients allow the calculation of their associated interfacial stiffnesses $K_L$, which are a direct measure of the adhesion at the interface; a higher $K_L$ corresponds to a stronger adhesion (see S.XI. for more information). 
A value of $K_L$ = 2.30 $\times$ 10$^{18}$  N/m$^3$ is found for the annealed flakes, which is larger than that of the non-annealed flakes, $K_L$ = 1.33 $\times$ 10$^{18}$  N/m$^3$. After annealing, the interfacial stiffness increases, resulting in a better transmission of the acoustic energy to the \ch{SiO2}/Si substrate during the successive reflections of the acoustic waves inside the Au/Cr/STO assembly, and therefore to a weaker reflection coefficient and a faster decay in amplitude.

Figure \ref{fig:pump_probe}d shows the Fourier transform of the temporal data in Fig. \ref{fig:pump_probe}c revealing that they are composed of 2 distinct frequencies. The low frequency ($\sim$22 GHz) corresponds to the first mode of standing waves in the Au/Cr/STO assembly unbound from the underlying \ch{SiO2}, calculated theoretically (see S.XI for details) at $f_{1U}$ = 21 GHz. The higher frequency ($\sim$57 GHz) likely corresponds to the third mode of the standing waves in the bounded assembly (calculated $f_{3B}$ = 53 GHz). The difference between the theoretical values and the experimental ones could be caused by small variations in the thickness of the Au/Cr/STO assembly and by the value of the longitudinal sound velocities used to calculate these frequencies taken from the literature (see S.XI). 
In both samples, the presence of frequencies corresponding to unbounded ($f_{1U}$) and bounded ($f_{3B}$) cases shows that the adhesion here is intermediate \cite{greener2019high} (between perfect contact and total debonding).
However, the amplitude of the components in the annealed case are weaker than in the non-annealed case (Fig. \ref{fig:pump_probe}d), since their attenuation by transmission of the acoustic energy to the substrate through multiple reflections in the Au/Cr/STO assembly is higher (difference in $\tau_{ac}$, Fig. \ref{fig:pump_probe}c). 
The increased adhesion found from these ultra-fast picosecond ultrasonics measurements is consistent with the increased hermeticity observed in Fig. \ref{fig:anneal}.



\begin{table*}[]
\centering
\small
\begin{tabular}{|c| >{\centering\arraybackslash}p{3cm}|>{\centering\arraybackslash}p{3cm}| >{\centering\arraybackslash}p{3cm} |>{\centering\arraybackslash}p{3cm}|}
\hline
                                    & \textbf{$\tau_{ac}$ (ps)}   & \textbf{$|R_{ac}|$}    & \textbf{$K_{L}$ (10$^{18}$ N/m$^3$)}    \\ \hline \hline

\textbf{Non-annealed (4 flakes)} &  219.9 $\pm$ 50.0  & 0.81 $\pm$ 0.04        & 1.33 $\pm$ 0.20        \\ \hline
\textbf{Annealed (5 flakes) }    &  113.6 $\pm$ 17.1    & 0.70 $\pm$ 0.04      & 2.30 $\pm$ 0.64  \\ \hline
\textbf{Theoretical values}   & \makecell{59 (perfect contact)\\ $\infty$ (total debonding)} & \makecell{0.45 (perfect contact)\\ 1 (total debonding)} & \makecell{\textgreater{}20 (perfect contact) \\ \textless{}0.1 (total debonding)} \\ \hline
\end{tabular}
\caption{Results of picosecond ultrasonics measurements on 4 non-annealed and 5 annealed STO samples.}
\label{table:results}
\end{table*}

%
%
%


\begin{figure}[htb!]
\includegraphics[width=\columnwidth]{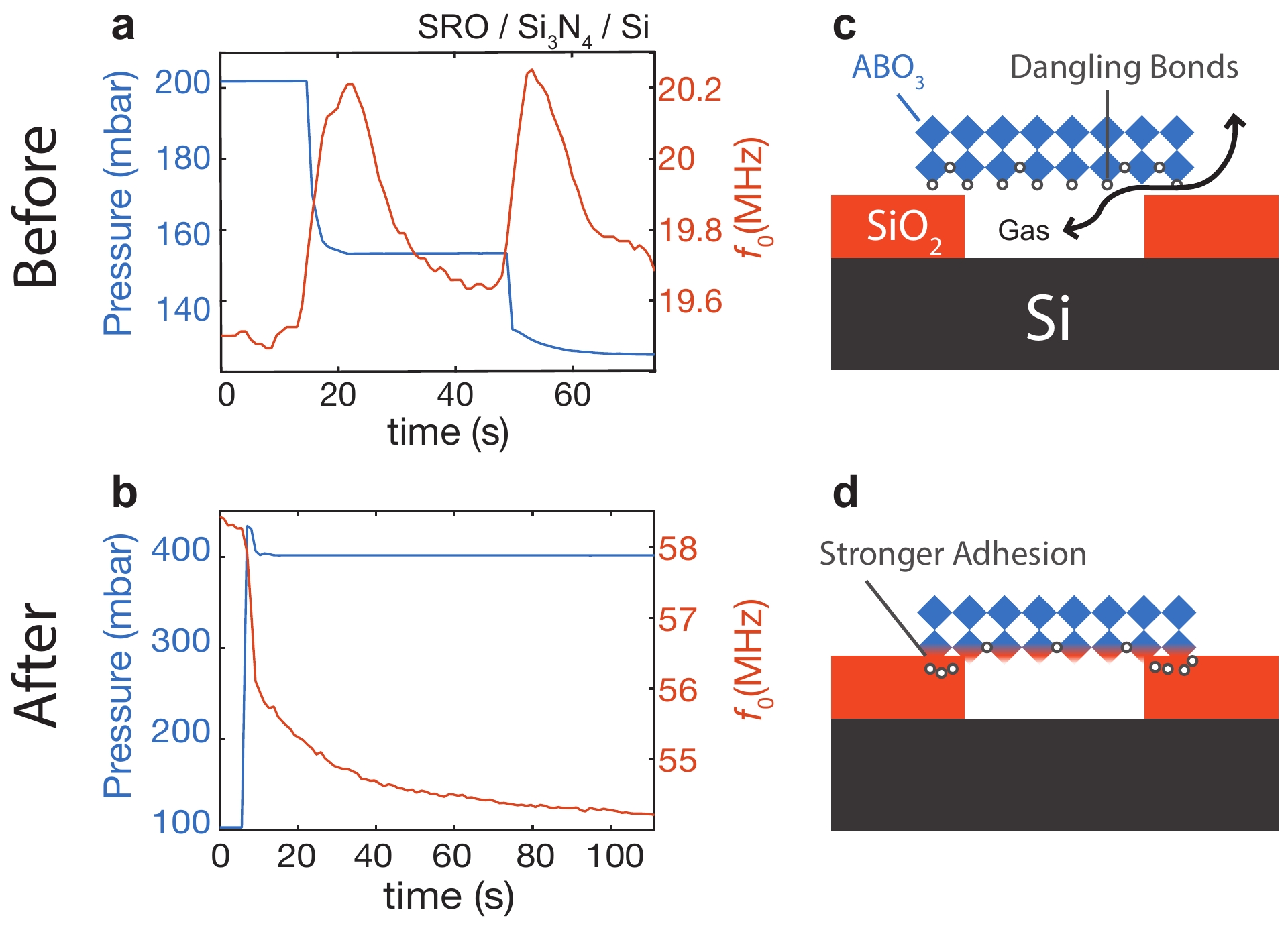}
\caption{Pressure response (left y-axis, blue) of mechanical resonance (right y-axis, orange) in a SRO device fabricated on 350 nm \ch{Si3N4/Si} \textbf{a} before annealing, and \textbf{b} after annealing.
Possible mechanism of the bonding is illustrated in \textbf{c\&d}. \textbf{c} Before annealing, there are dangling bonds at the bottom of the flakes. The vdW gap between the SRO and the \ch{SiO2} allows for gases to pass through. \textbf{d} After annealing, vacancies bond with the oxygen in the substrate leading to a stronger bond to form at the interface.}
\label{fig:mechanism}
\end{figure}

Having used mechanical resonance measurements and picosecond ultrasonics measurements to establish the enhancement of the adhesion at the interface of complex oxides and \ch{SiO2}, we now move on to investigate whether the observed adhesion increase is unique to the \ch{SiO2} substrate. For this purpose, we performed additional permeation measurements on nominally identical samples made on an oxygen-free substrate, \ch{Si3N4}/Si. Figures \ref{fig:mechanism}a\&b, show the time dependence of the resonance frequency of a SRO flake suspended over a cavity etched in \ch{Si3N4}/Si. Figure \ref{fig:mechanism}a is taken before annealing and Fig. \ref{fig:mechanism}b is taken after annealing for 1 hour at 400 $^\circ$C. Before annealing, the permeation time constant is $\tau_p$ = 6.02 s and after annealing for 1 hour, it increases to $\tau_p$ = 22.5 s. Although a factor of 3.7 improvement is observed, the absolute leakage time constant after annealing in \ch{Si3N4} devices are on the order of those in \ch{SiO2} devices even before annealing (see  S.V for the analysis on $\tau_p$). Investigation by means of energy dispersive X-ray spectroscopy (EDX) ruled out any role of SAO residues in the adhesion (see S.VI). The absence of the Al peak in the EDX spectra strongly suggests that SAO is indeed fully removed by water. Due to above reasons we hypothesize that the improved bonding is mediated by the presence of oxygen in the \ch{SiO2} substrate. A possible scenario is illustrated in Fig. \ref{fig:mechanism}c-d where the adhesion is enhanced by the reaction between dangling bonds at the bottom of the complex oxide flake and the oxygen rich substrate at elevated temperatures.


Furthermore, it is worth to note the longevity of the improved adhesion on top of the \ch{SiO2} substrate. We have performed pick-up techniques widely used in the fabrication of vdW heterostructures using both polypropylene carbonate (PPC) and polycarbonate (PC), the latter of which has stronger adhesive properties and is thus more suitable for monolayer transfer \cite{pizzocchero2016hot,zomer2014fast, caneva2020mechanically}. Non-annealed samples of SRO on \ch{SiO2} were easily picked up using both PPC and PC while the samples annealed 8 months ago and stored in ambient conditions, could not be removed from the substrate (see S.VIII). Unlike vdW materials which can be detached from the substrate after annealing and subsequently storing in ambient conditions, the improvement in the adhesion seems to be longer-lasting in annealed oxides.


%
%

\section{Discussion \& Conclusion}

Although graphene and its family of vdW materials have demonstrated superior pressure sensing capabilities \cite{lemme2020nanoelectromechanical} compared to the state-of-the-art made from Si \cite{vsivskins2020sensitive}, the leakage through the vdW gap between the material and the substrate still remains a key challenge to overcome. 
Previous works have shown that the hermeticity can be improved by e.g. ironing with a diamond atomic force microscopy (AFM) tip \cite{manzanares2020improved} or electron-beam induced deposition of \ch{SiO2} \cite{lee2019sealing}. Unfortunately, both the AFM and EBID methods are not scalable, since they are too slow to apply over large areas. Furthermore, if only the edge of the flake is sealed, and there is a puncture in a sealed flake, then all of the cavities underneath the flake are effectively vented. The intrinsic vdW nature and the difficulty in producing pin-hole free 2D materials are the hurdles in fabricating reliable hermetically sealed pressure sensors.

The annealing procedure performed on the single crystal complex oxide flakes on \ch{SiO2} improves the hermeticity of the cavity as measured by mechanics and improves mechanical contact as measured by picosecond ultrasonics. Furthermore, the PC pick-up technique widely used in the fabrication of the vdW heterostructures is ineffective in removing the annealed flakes from \ch{SiO2} even after 8 months of storage in ambient conditions. Therefore we can conclude that after the self-sealing procedure the adhesion is better and contact between the PLD layer and the substrate is more intimate. This increase can be caused by removal of water, and by formation of chemical bonds. Both mechanisms might play a role. However, the permeation time constant does not seem to increase by comparable magnitudes in samples created on \ch{Si3N4}/Si substrates. This seems to suggest that chemical bond formation is the most likely, as water removal will likely happen for both substrates. This mechanism is similar to direct wafer-to-wafer bonding techniques that are in widespread use in the semiconductor industry \cite{masteika2014review, schmidt1998wafer, alexe2013wafer, harendt1992silicon, tong1994semiconductor}.



Thanks to the advent of the water-releasing technique \citep{lu2016synthesis}, free-standing single-crystal complex oxides can be synthesized and transferred onto cavities for MEMS applications as presented in this work. One key difference between vdW materials and free-standing complex oxides is the existence of interlayer covalent bonds. We propose that by controlling the concentration of oxygen vacancies in the complex oxide, the density of dangling bonds at the surface can be tuned. 
Therefore, we expect to be able to further improve the adhesion - thus the hermeticity - by optimizing the annealing conditions. Yong \textit{et al.} have shown that at temperatures above 750 $^\circ$C and pressures below 400 mTorr, interfacial species of \ch{TiSi2} and/or \ch{SrSiO3} form at the STO-\ch{SiO2} interface \cite{yong2010thermal}. We believe similar effects may be happening in our flakes although our annealing temperatures and \ch{O2} partial pressures are quite different.

In summary, we have investigated the use of free-standing complex oxides \ch{SrRuO3} and \ch{SrTiO3} for pressure sensing applications and presented a self-sealing method based on annealing to improve the hermeticity of complex oxide based pressure sensors. Gases permeate along the vdW - substrate interface and the elimination of this leakage path is a key towards fabricating next generation pressure sensors. We realized a leap towards this goal by promoting stronger adhesion to form at the interface of complex oxides and \ch{SiO2}. Improvements in the gas permeation time constant  as well as the contrast in the acoustic impedance at the interface suggest that the interface adhesion of complex oxides (\ch{SrRuO3} and \ch{SrTiO3}), and \ch{SiO2} is stronger after annealing. We further investigated the effect of the substrate on the interfacial adhesion by performing the permeation measurements on devices made on \ch{Si3N4/Si}. Since significant improvements in time constant are only observed on \ch{SiO2} substrate, it is likely that the oxygen atoms on the \ch{SiO2} substrate surface play an important role in mediating the adhesion. Our work presents a first step towards implementing free-standing complex oxides as an alternative to silicon and 2D materials in next generation MEMS and NEMS sensors.






%
%
%
%
%
%

\begin{acknowledgments}

M.L., H.S.J.v.d.Z. and P.G.S. acknowledge funding from the European Union’s Horizon 2020 research and innovation program under grant agreement number 881603. 
A.D.C. acknowledges funding from Quantox of QuantERA ERA-NET Cofund in Quantum Technologies and by the Netherlands Organisation for Scientific Research (NWO/OCW) as part of the VIDI program.
G.V. acknowledges support from project TKI-HTSM/19.0172.
\end{acknowledgments}


%
%
%

\bibliography{selfsealing}
\bibliographystyle{ieeetr}
\end{document}


\renewcommand{\theequation}{S.\arabic{equation}}
\renewcommand{\thefigure}{S\arabic{figure}}
\renewcommand{\thetable}{S\arabic{table}}


\title{{\large Supplementary Information for: }\\ Self-sealing complex oxide resonators}
\author{Martin Lee}
\affiliation{Kavli Institute of Nanoscience, Delft University of Technology, Lorentzweg 1, 2628 CJ Delft, The Netherlands.}

\author{Martin Robin}
\affiliation{Department of Precision and Microsystems Engineering, Delft University of Technology, Mekelweg 2, 2628 CD Delft, The Netherlands.}

\author{Ruben Guis}
\affiliation{Department of Precision and Microsystems Engineering, Delft University of Technology, Mekelweg 2, 2628 CD Delft, The Netherlands.}

\author{Ulderico Filippozzi}
\affiliation{Kavli Institute of Nanoscience, Delft University of Technology, Lorentzweg 1, 2628 CJ Delft, The Netherlands.}

\author{Dong Hoon Shin}
\affiliation{Kavli Institute of Nanoscience, Delft University of Technology, Lorentzweg 1, 2628 CJ Delft, The Netherlands.}

\author{Thierry C. van Thiel}
\affiliation{Kavli Institute of Nanoscience, Delft University of Technology, Lorentzweg 1, 2628 CJ Delft, The Netherlands.}

\author{Stijn Paardekooper}
\affiliation{Department of Precision and Microsystems Engineering, Delft University of Technology, Mekelweg 2, 2628 CD Delft, The Netherlands.}

\author{Johannes R. Renshof}
\affiliation{Kavli Institute of Nanoscience, Delft University of Technology, Lorentzweg 1, 2628 CJ Delft, The Netherlands.}

\author{Herre S. J. van der Zant}
\affiliation{Kavli Institute of Nanoscience, Delft University of Technology, Lorentzweg 1, 2628 CJ Delft, The Netherlands.}

\author{Andrea D. Caviglia}
\affiliation{Kavli Institute of Nanoscience, Delft University of Technology, Lorentzweg 1, 2628 CJ Delft, The Netherlands.}

\author{Gerard J. Verbiest}
\affiliation{Department of Precision and Microsystems Engineering, Delft University of Technology, Mekelweg 2, 2628 CD Delft, The Netherlands.}

\author{Peter G. Steeneken}
\affiliation{Kavli Institute of Nanoscience, Delft University of Technology, Lorentzweg 1, 2628 CJ Delft, The Netherlands.}
\affiliation{Department of Precision and Microsystems Engineering, Delft University of Technology, Mekelweg 2, 2628 CD Delft, The Netherlands.}

\maketitle
%
%
%
\pagebreak


\section{Methods}
\label{supp:methods}

\subsection*{Pulsed laser deposition of SRO/SAO/STO}
The SRO and SAO films are grown by pulsed laser deposition on \ch{TiO2} termiated STO(001) substrates purchased from CrysTec GmbH. Pulses of KrF excimer laser are delivered at 1 Hz at fluences of 1.7 J/cm$^{2}$. SAO is grown at 800 $^\circ C$ and 2$\times$10$^{-6}$ mbar pressure and SRO at 550 $^\circ C$ and 1$\times$10$^{-1}$ mbar \ch{O2} pressure. After growth, the stack is annealed at 500 $^\circ C$ for an hour and cooled down to room temperature in 300 mbar \ch{O2}.

\subsection*{Pulsed laser deposition of STO/SAO/STO}
The freestanding STO film is fabricated in a similar manner to the SRO. STO is grown in 1$\times$10$^{-6}$ mbar \ch{O2} pressure at 800 $^\circ C$. The excimer laser is delivered at 1 Hz at a fluence of 1.2 J/cm$^2$. After growth, the stack is annealed at 500 $^\circ C$ for an hour and cooled down to room temperature in 300 mbar \ch{O2}.

\subsection*{Release and Transfer onto bare \ch{SiO2/Si}}

Prior to releasing from the growth substrate, a commercial polydimethylsiloxane (PDMS) film (Gel-Pak \textregistered) is attached to the surface of the SRO/SAO/STO stack. The stack covered by PDMS is then submerged under deionized water for 24 hours. Once the SAO is fully etched away and only SRO remains on the PDMS, the substrate detaches from the PDMS. The SRO/PDMS stamp is removed from water, dried in with \ch{N2} gas and is ready to stamp. The flakes are then transferred onto a bare \ch{SiO2/Si} chip using the viscoelastic stamping technique \cite{castellanos2014deterministic}. This dummy \ch{SiO2/Si} chip is used as an intermediate substrate to place down the flakes for XRD characterization. From this dummy chip, flakes of adequate size are chosen for the final transfer using an optical microscope. 

\subsection*{Polymer assisted pick up technique}

The polymer assisted pickup technique using polypropylene carbonate (PPC) \cite{pizzocchero2016hot} coated PDMS dome \cite{kim2016van} is used to transfer the chosen flake from the dummy \ch{SiO2/Si} to the device chip. The PPC covered PDMS dome is brought in contact with the chosen flake and the chip is heated beyond the $T_g \ \sim $ 45 $^\circ$C. Once PPC is molded into the shape of the flake, the stack is cooled down to room temperature and the dome along with the flake is detached from the \ch{SiO2/Si}. The flake is then transferred onto a prepatterned \ch{SiO2/Si} device consisting cavities etched into \ch{SiO2}/Si. Finally the flake is released above 60 $^\circ$C leaving only the flake.

\subsection*{Prepatterend \ch{SiO2/Si}}

Dry thermal oxide of 285 nm, grown on highly doped (Si++) silicon is used as the substrate. Vistec EBPG 5000+ is used to expose the cavity defined in 500 nm of AR-P 6200 positive e-beam resist. After exposure and development, the cavities are dry etched into the \ch{SiO2/Si} using \ch{CHF3} and Ar plasma until all the \ch{SiO2} is removed. Remainder of AR-P 6200 is removed in PRS-3000 over night, rinsed and blow-dried. The substrates are further cleaned in \ch{O2} plasma asher for 3 minutes. Similar procedure is employed for \ch{Si3N4}/Si chips with 350 nm LPCVD grown \ch{Si3N4}.

\subsection*{Annealing}
The annealing takes place in ambient conditions on top of a VWR hot plate above 300 $^\circ$C for 15 minutes to 1 hr.

\subsection*{Samples for ultra-fast acoustics}

For the measurements of the acoustic boundary conditions using ultra-fast pump-probe method, separate samples need to be fabricated with metal layers deposited on top of the flakes. The metal layer is necessary to reflect the probe and absorb the pump to generate an acoustic pulse.
Two dummy \ch{SiO2/Si} are prepared with several flakes of 82 nm STO. On both chips, 3 nm of Cr and 30 nm of Au are deposited using a Temescal e-beam evaporator.

\clearpage

\section{RHEED}

\begin{figure}[ht]
\includegraphics[width=\columnwidth]{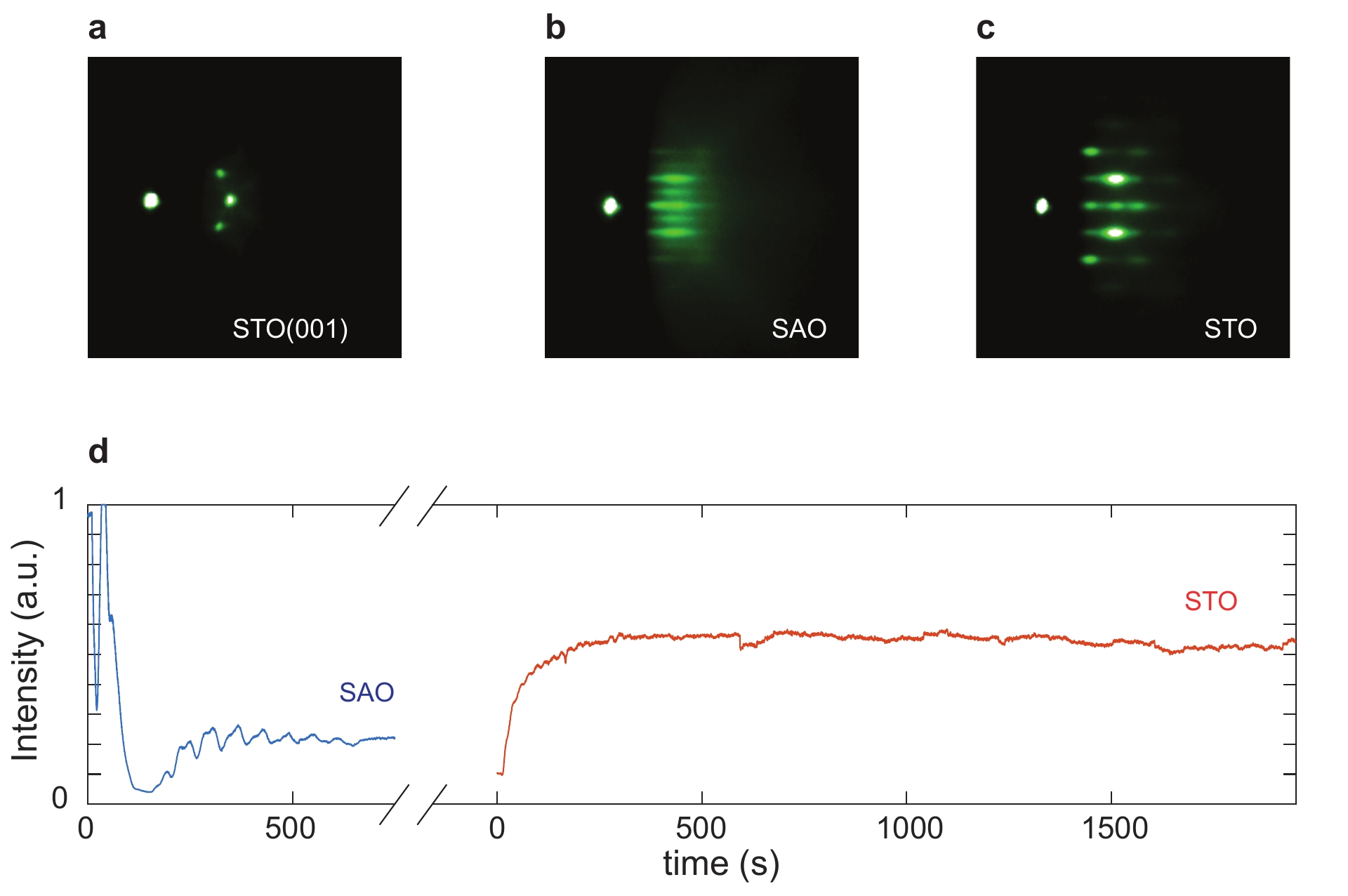}
\caption{Reflection high energy electron diffraction images of \textbf{a} the STO(001) substrate, \textbf{b} SAO buffer layer on STO(001), and \textbf{c} STO on SAO/STO(001). \textbf{d} RHEED oscillations of SAO and STO monitored during growth}
\label{suppfig:rheed_sto}
\end{figure}

Reflection high energy electron diffraction (RHEED) is used to monitor the growth of the 82 nm STO used in the main text. In Fig. \ref{suppfig:rheed_sto}a-c, RHEED images of the substrate, SAO buffer layer and STO (82 nm) are shown. The RHEED intensities of the side spots/streaks are used to monitor the layer-by-layer growth of SAO and STO as shown in Fig. \ref{suppfig:rheed_sto}d. 

\clearpage
\section{Characterization of STO}

\begin{figure}[ht]
\includegraphics[width=\columnwidth]{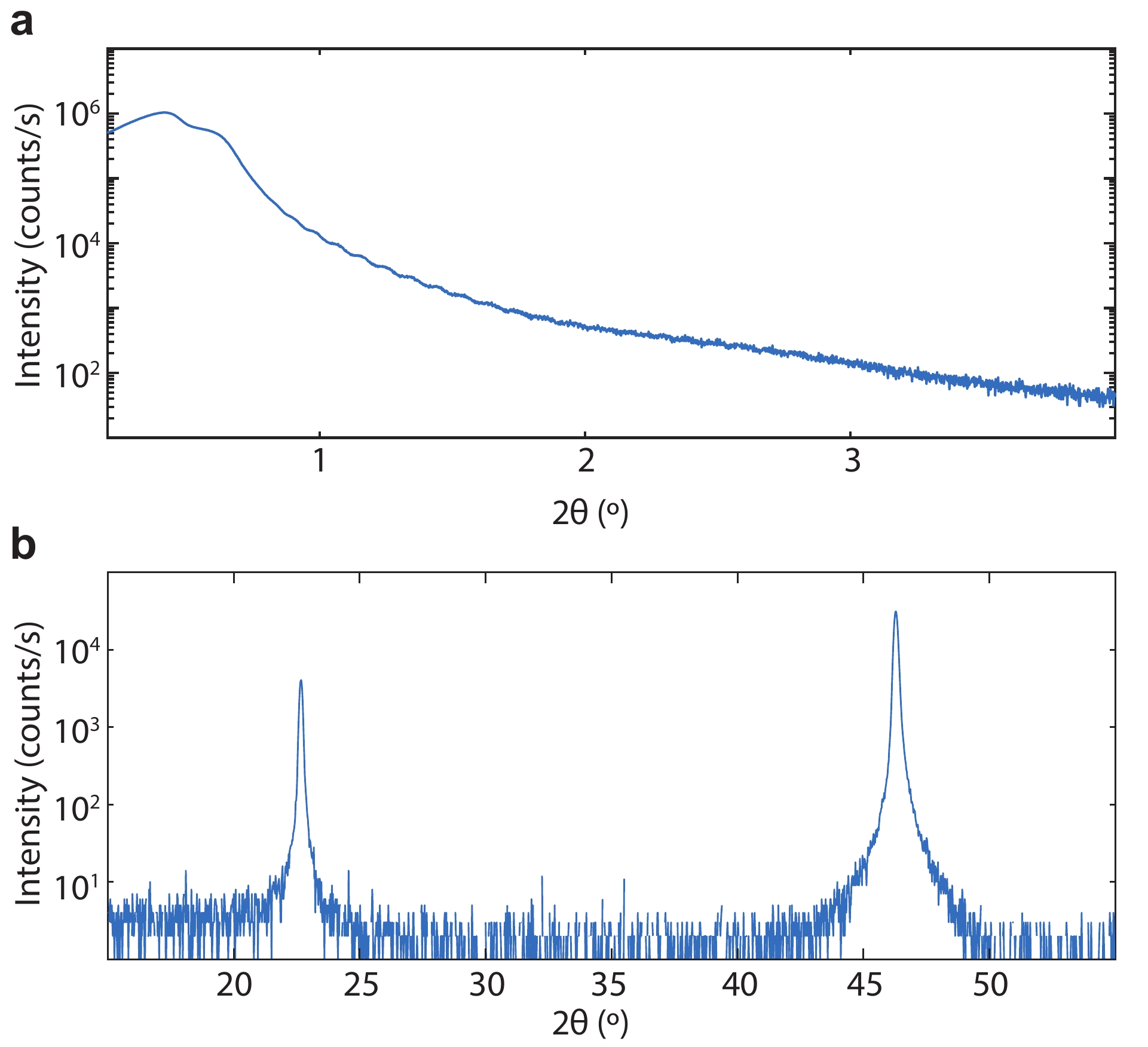}
\caption{\textbf{a} X-ray refractometry (XRR) performed on free-standing STO stamped on \ch{SiO2/Si}. \textbf{b} X-ray diffraction (XRD) of the same film showing STO (001) and STO (002) peaks.}
\label{suppfig:xrd_xrr_sto}
\end{figure}

\begin{figure}[ht]
\includegraphics[width=\columnwidth]{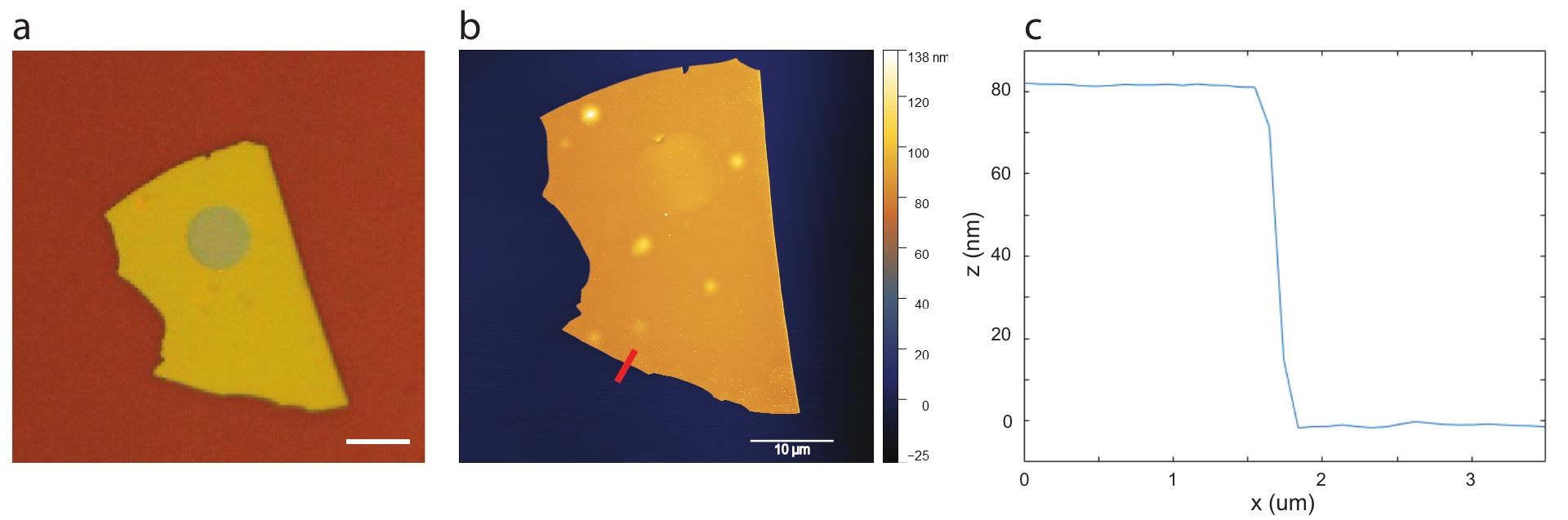}
\caption{\textbf{a} Optical image of a STO flake suspended over a 10 \textmu m cavity. Scalebar: 10 \textmu m. \textbf{b} Atomic force microscopy (AFM) image of the device. Red line indicates where a profile is taken. \textbf{c} Profile at the edge of the flake as indicated by the red line in \textbf{b}. Thickness of 82 nm is measured.}
\label{suppfig:afm_sto}
\end{figure}

Figure \ref{suppfig:xrd_xrr_sto}a shows a X-ray refractometry (XRR) performed on a 218 nm thick STO stamped on \ch{SiO2/Si} and Figure \ref{suppfig:xrd_xrr_sto}b shows a X-ray diffraction (XRD) of the same film. The XRR shows multiple oscillations occuring between 1 $^\circ$ and 2 $^\circ$ indicating the uniformity of the film. Fast fourier transform analysis of this film gives a peak near 85 nm which agrees well with the atomic force microscopy (AFM) data shown in Fig. \ref{suppfig:afm_sto}

\clearpage
\section{Permeation time constants}

\subsection*{SRO device} 

\begin{figure}[ht!]
\includegraphics[width=0.6\columnwidth]{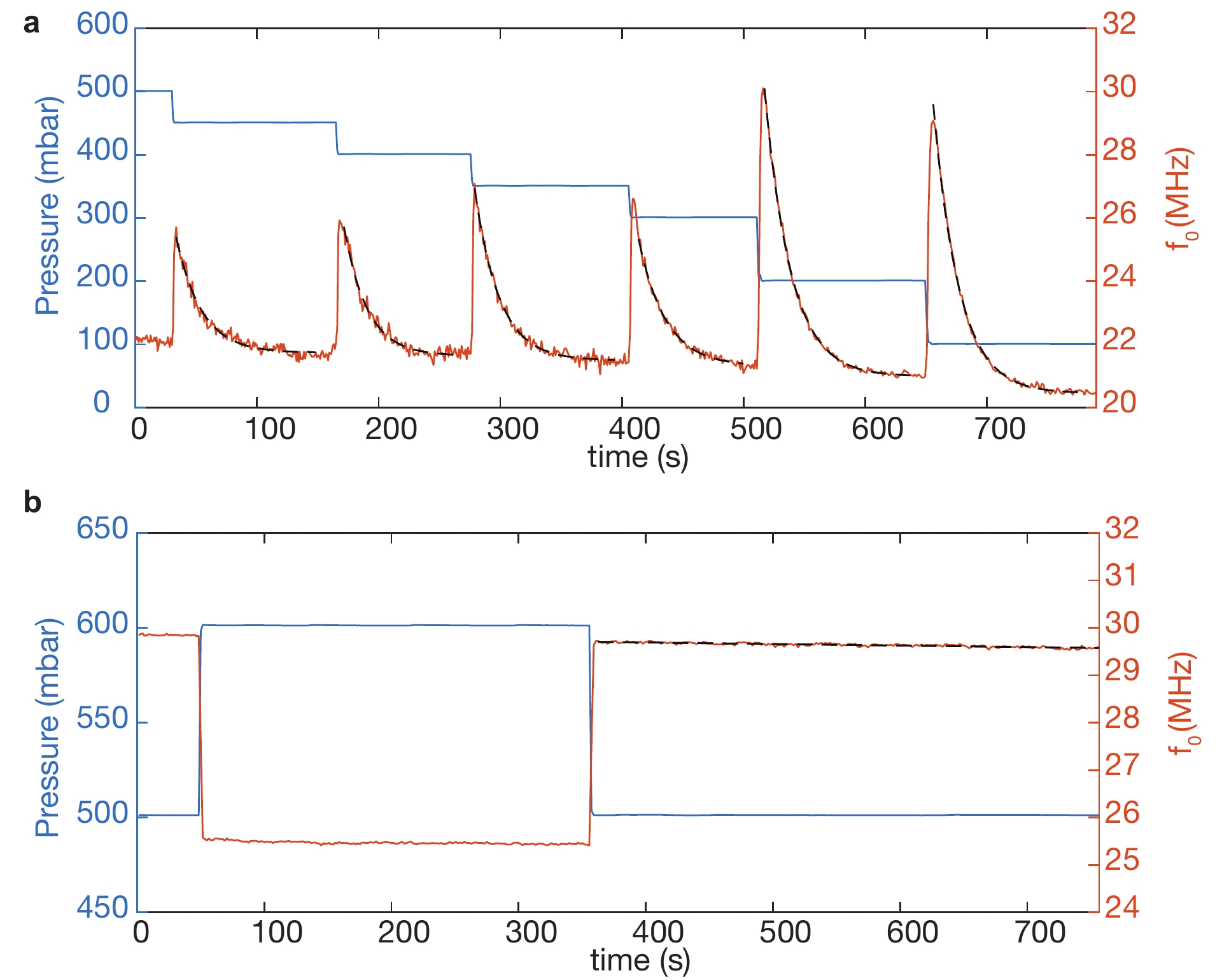}
\caption{Pressure (left y-axis, blue) response of mechanical resonance (right y-axis, orange) of SRO \textbf{a} before annealing \textbf{b} and after annealing. Black dashed lines are the decay curves taken to extract the permeation time constant $\tau_p$. }
\label{suppfig:SRO}
\end{figure}

Figure \ref{suppfig:SRO} show the pressure response of the resonance frequency before and after annealing. Before annealing, an average permeation time constant of $\tau_p $ = 20.75 seconds is extracted. Individual $\tau_p$ are listed in table \ref{supp:table:tau:SRO}. After annealing, this increases to $\tau_p$ = 1.1$\times$10$^4$ seconds.

\begin{table}
\centering
\footnotesize
\begin{tabular}{l|r}
&$\tau_p $(s)\\ \hline
curve 1& 20.79  \\
curve 2& 18.95 \\
curve 3& 20.14 \\
curve 4& 22.16 \\
curve 5& 21.45 \\
curve 6& 21.02\\ \hline
mean & 20.75
\end{tabular}
\caption{Permeation time constants of decay curves in Fig. \ref{suppfig:SRO}a}
\label{supp:table:tau:SRO}
\end{table}

\subsection*{STO device} 

\begin{figure}[ht]
\includegraphics[width=0.6\columnwidth]{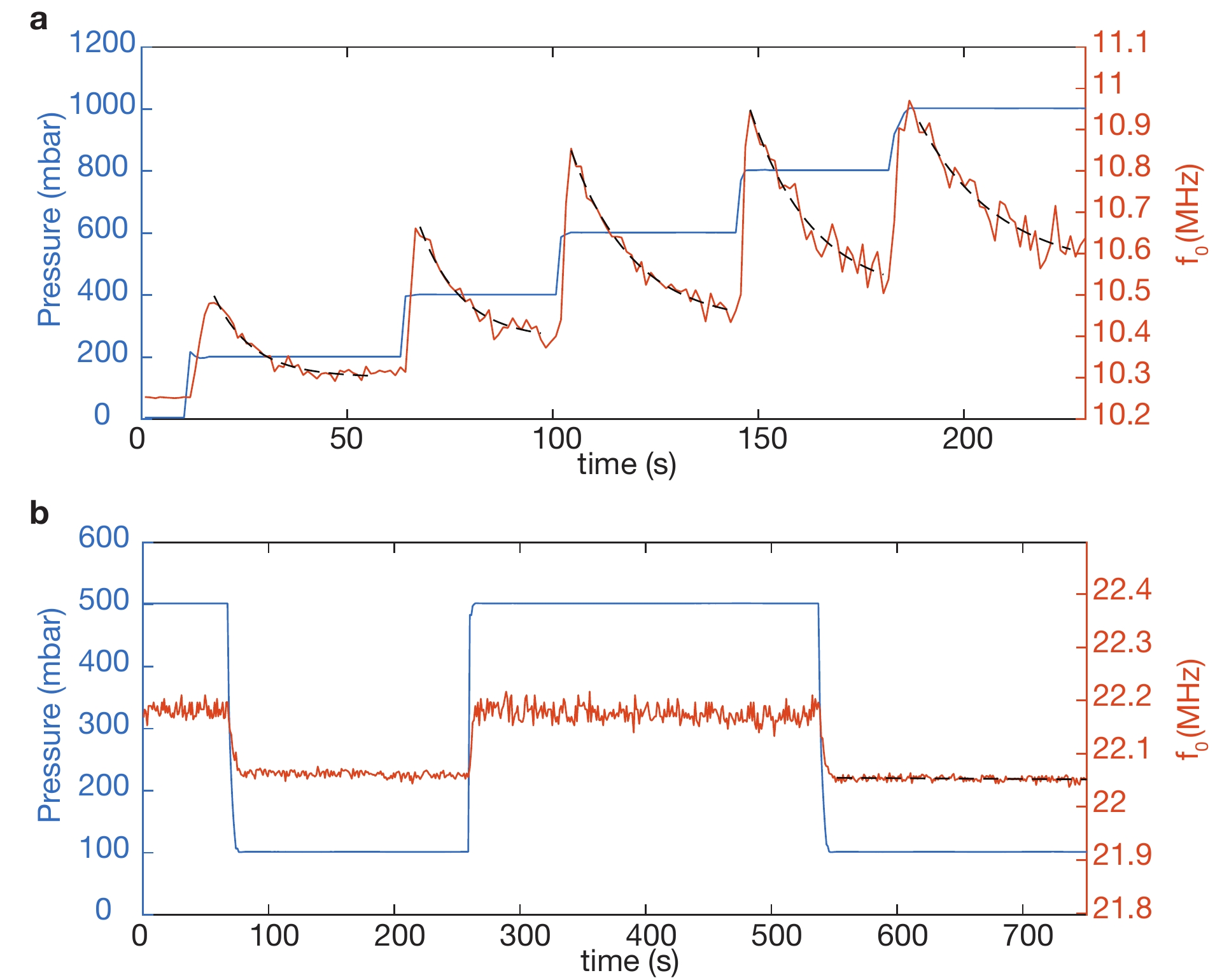}
\caption{Pressure (left y-axis, blue) response of mechanical resonance (right y-axis, orange) of STO \textbf{a} before annealing \textbf{b} and after annealing. Black dashed lines are the decay curves taken to extract the permeation time constant $\tau_p$. }
\label{suppfig:STO}
\end{figure}

Figure \ref{suppfig:STO} show the pressure response of the resonance frequency before and after annealing. Before annealing, an average permeation time constant of $\tau_p $ = 13.87 seconds is extracted. Individual $\tau_p$ are listed in table \ref{supp:table:tau:STO}. After annealing, this increases to $\tau_p$ = 1.2$\times$10$^5$ seconds.

\begin{table}
\centering
\footnotesize
\begin{tabular}{l|r}
&$\tau_p $(s)\\ \hline
curve 1& 9.34  \\
curve 2& 10.16 \\
curve 3& 15.24 \\
curve 4& 15.17 \\
curve 5& 19.46 \\ \hline
mean & 13.87
\end{tabular}
\caption{Permeation time constants of decay curves in Fig. \ref{suppfig:STO}}
\label{supp:table:tau:STO}
\end{table}

\clearpage
\section{SRO on \ch{Si3N4}} 

\begin{figure}[ht]
\includegraphics[width=0.5\columnwidth]{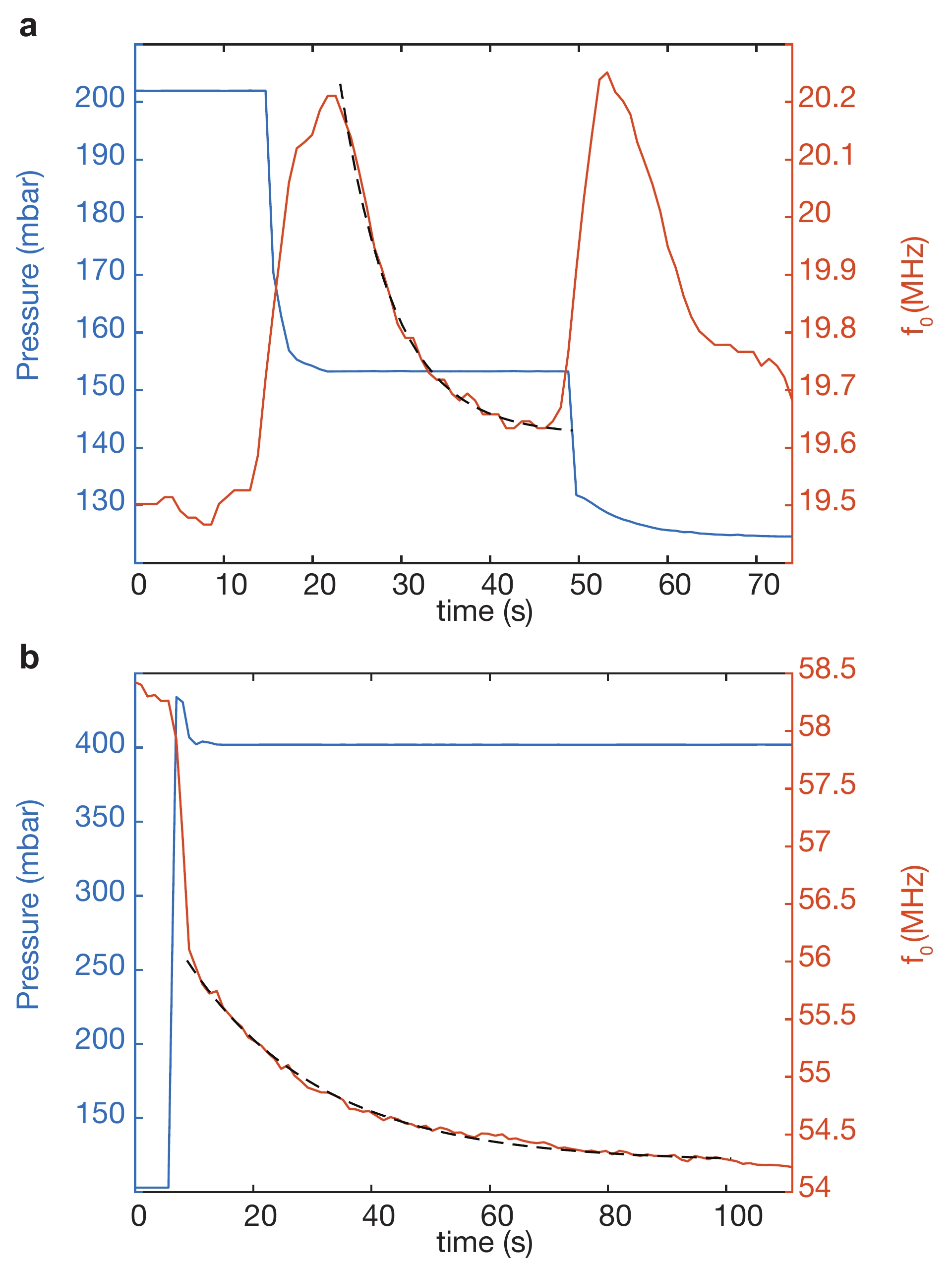}
\caption{Pressure response (left y-axis, blue) of mechanical resonance (right y-axis, orange) in a SRO device fabricated on 350 nm \ch{Si3N4/Si} \textbf{a} before annealing, and \textbf{b} after annealing. Dashed black lines are exponential fits to extract the $\tau_p$. Before annealing, $\tau_p$ = 6.02 s and after annealing $\tau_p$ = 22.5 s.}
\label{suppfig:SIN}
\end{figure}

Figure \ref{suppfig:SIN}a\&b are the pressure dependent resonance frequency data taken on a SRO flake stamped on \ch{Si3N4} substrate. A permeation time constant of $\tau_p$ = 6.02 s is extracted from the dashed black line in the data before annealing in Fig. \ref{suppfig:SIN}a. Similarly, a $\tau_p$ = 22.5 s is extracted from the dashed black line in the data after annealing in Fig. \ref{suppfig:SIN}b.

\clearpage
\section{EDX of free-standing SRO}

It is worth noting that the atomic force microscopy (AFM) profile of SRO flakes on \ch{SiO2/Si} are 10.9 nm while the XRD data shown in the main text suggests a thickness of 6.3 nm. The discrepancy in the measured thicknesses between the two methods can be attributed to a few things. First, there could be an overestimation of the thickness from AFM measurements due to tenting effect, water layer, or dirt. Shearer \textit{et al.} have shown that the van der Waals gap between a flake and the substrate can often give a gross overestimation on the thickness of the flake when measured with an AFM \cite{shearer2016accurate}. Second, SAO may not have fully etched away, leaving behind some pillars of residues which also can attribute to the tenting effect. It is possible to hypothesize that the bonding is mediated by the residual SAO layer where the half etched SAO residues provide dangling bonds which could readily bind with the \ch{SiO2} underneath. In order to eliminate the hypothesis that this bonding is mediated by the remnant SAO layers providing reactive dangling bonds underneath the SRO and STO flakes, we have performed energy dispersive X-ray spectroscopy (EDX) of the flakes and AFM topography analysis of the STO substrate after the release of the layers. Due to the absence of the Al peak in the EDX as well as the stepping terraces observed in the AFM of the substrate (next section), we can safely rule out the hypothesis that there is a remnant SAO residue which bonds to the \ch{SiO2}.

\begin{figure}[ht]
\includegraphics[width=1\columnwidth]{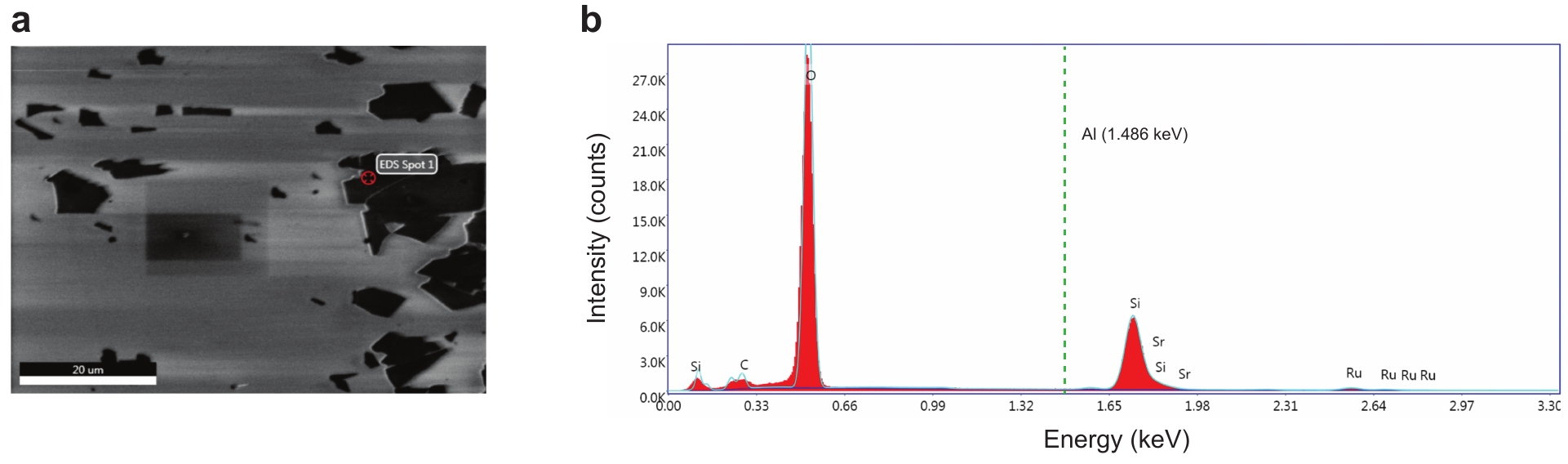}
\caption{\textbf{a} Scanning electron microscopy image of free-standing SRO stamped on \ch{SiO2/Si} taken at 52 degree tilt . The spot where energy dispersive X-ray spectroscopy (EDX) is performed is indicated with a marker. \textbf{b} EDX of the film showing contributions from Si, O, Sr and Ru but not Al (1.486 keV green dashed line) suggesting that \ch{Sr3Al2O6} is fully etched away in water.}
\label{suppfig:SEM_EDX}
\end{figure}

Figure \ref{suppfig:SEM_EDX}a shows a scanning electron microscopy image (SEM) of the SRO flakes stamped on \ch{SiO2/Si}. An EDX spectrum is taken in one of the flakes indicated by a marker in Fig. \ref{suppfig:SEM_EDX}a. The EDX spectrum plotted in Fig. \ref{suppfig:SEM_EDX}b shows the elemental analysis which includes Si, O, Sr and Ru. These arise from both the flake and the substrate. However, there is no peak at 1.486 keV (green dashed line) which corresponds to the energy of Al. This indicates that the sacrificial water soluble SAO layers underneath the films of interest (SRO and STO) are fully etched away. SEM and EDX are performed using FEI Helios G4 CX at 5 keV and at a tilt of 52 degrees.

\clearpage
\section{AFM of substrate after water etching}

\begin{figure}[ht]
\includegraphics[width=0.8\columnwidth]{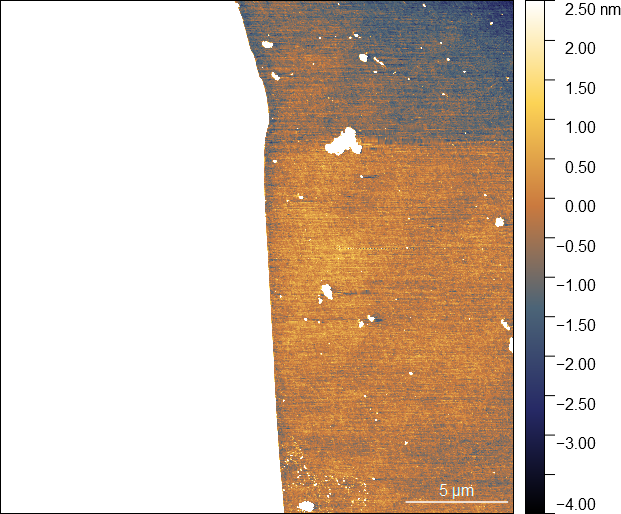}
\caption{AFM image of the substrate after water etching. Bright left side still has SRO/SAO on STO(001) and the right side is the bare STO(001) surface after SAO is etched away. Signs of terraces are visible.}
\label{suppfig:afm_substrate}
\end{figure}

Figure \ref{suppfig:afm_substrate} shows an AFM image of the substrate after water etching. The bright left side still has SRO/SAO while the right side is the bare STO (001) substrate. Hints of terraces can still be seen on the bare STO after etching away the SAO, suggesting that most of the SAO is removed in water. 

\clearpage
\section{Optical images and AFM of flakes not being picked up by polymer}

\begin{figure}[ht]
\includegraphics[width=1\columnwidth]{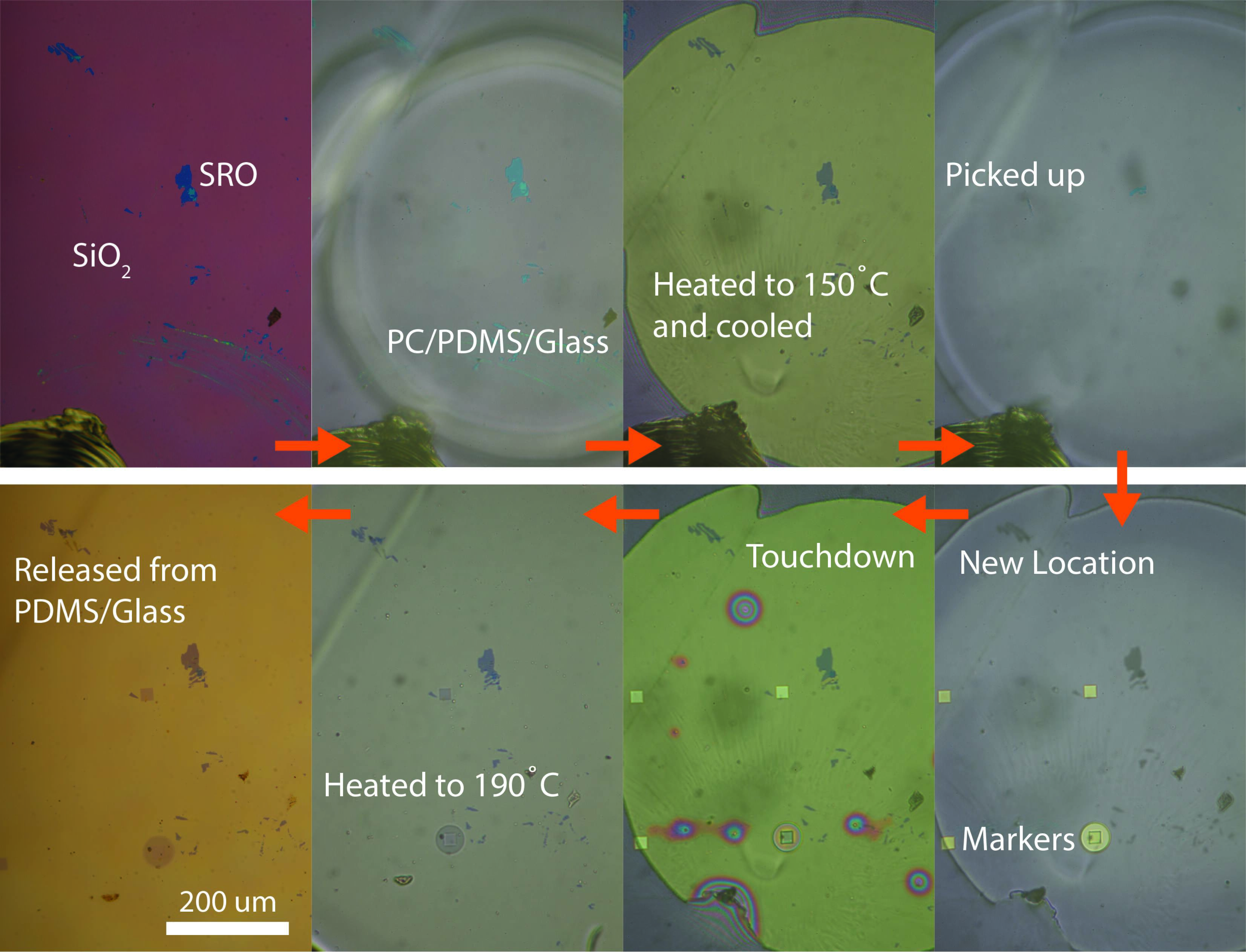}
\caption{Optical microscopy images of the flake pick-up process showing the transfer of non-annealed SRO from a dummy \ch{SiO2/Si} substrate to a patterned substrate. Polycarbonate is used as a sticky polymer to transfer the flakes (in this case, SRO).}
\label{suppfig:fresh_pickup}
\end{figure}

\begin{figure}[ht]
\includegraphics[width=1\columnwidth]{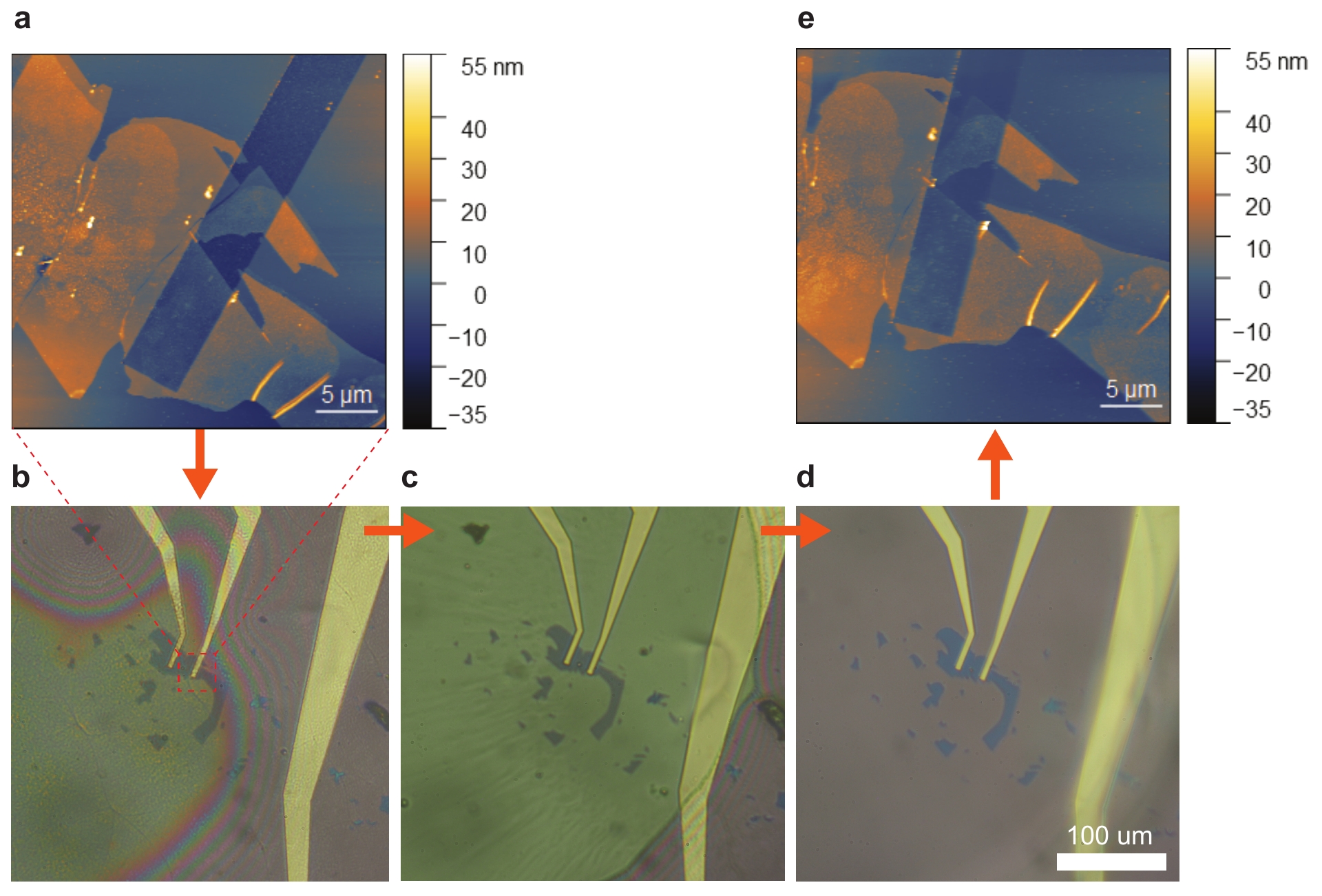}
\caption{AFM and optical microscopy images highlighting the inability to pick up the flakes of annealed SRO on \ch{SiO2/Si}. Polycarbonate is used as a sticky polymer to transfer the flakes. \textbf{a} AFM image of an annealed flake on top of a recessed electrode. \textbf{b-d} Optical images of the attempt to pick up the annealed flake using polycarbonate/ PDMS. \textbf{e} AFM image of the same same area after the pickup attempt.}
\label{suppfig:not_pickup}
\end{figure}

In this section, we provide an additional evidence for the improved adhesion of complex oxide flakes on \ch{SiO2} upon annealing. In Figs. \ref{suppfig:fresh_pickup} and \ref{suppfig:not_pickup}, we compare the maneuverability of non-annealed and annealed SRO flakes, respectively. In Fig. \ref{suppfig:fresh_pickup}, SRO flake on a dummy \ch{SiO2} is picked up using a standard polycarbonate/PDMS dome technique \cite{zomer2014fast,kim2016van}. It can be observed that the non-annealed SRO flakes are easily picked up from the original dummy \ch{SiO2} and transferred to a patterned \ch{SiO2}/Si. However, the same method used on annealed SRO flakes is unable to detach the flakes from the \ch{SiO2/Si} substrate. Figure \ref{suppfig:not_pickup} illustrates the process of the attempt to pick up annealed SRO flakes. In Fig. \ref{suppfig:not_pickup}a, an AFM scan is shown of an area consisting of a flake stamped on top of a Pd electrode recessed into \ch{SiO2/Si}. In Fig. \ref{suppfig:not_pickup}b-d, optical images of the pick up process attempt is shown. In Fig. \ref{suppfig:not_pickup}d, it can already be seen that the SRO flake remains intact, seemingly unaffected by the transfer attempt. In Fig. \ref{suppfig:not_pickup}e an AFM scan of the same area as Fig. \ref{suppfig:not_pickup}a after the pick up attempt is shown. It shows that the flake is not damaged by the attempt. It is worth mentioning that we have not succeeded in picking up SRO from \ch{SiO2} after annealing, but using a sharp needle in a probe station, we were able to scratch away SRO from \ch{SiO2}.

\clearpage
\section{AFM of flakes}

\begin{figure}[ht]
\includegraphics[width=0.5\columnwidth]{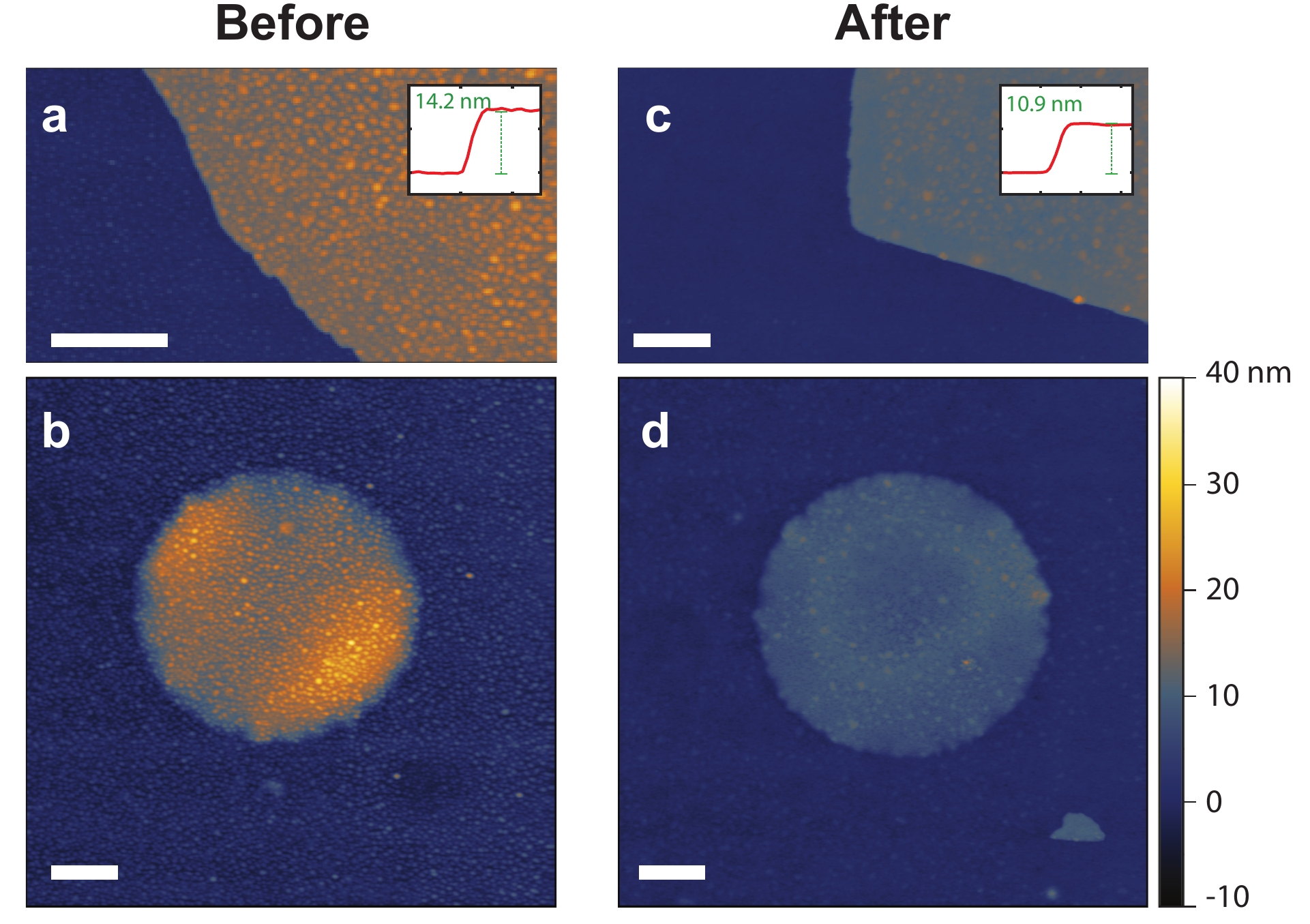}
\caption{\textbf{a} Atomic force microscopy (AFM) image of the edge of the SRO flake before annealing. The inset shows the step height of 14.2 nm. \textbf{b} AFM image of the suspended drum before annealing. \textbf{c} AFM image of the edge of the SRO flake after annealing. The inset shows the step height of 10.9 nm. \textbf{d} AFM image of the suspended drum after annealing.}
\label{suppfig:AFM}
\end{figure}

In this section we use atomic force microscopy (AFM) to inspect SRO flakes before and after annealing. Figure \ref{suppfig:AFM}a\&b show the AFM images of SRO flakes across the edge of the flake (Fig. \ref{suppfig:AFM}a) and on the drum (Fig. \ref{suppfig:AFM}b). The line scan over the edge of the flake shows a step height of 14.2 nm (inset of Fig. \ref{suppfig:AFM}a). Compared to the XRD data shown in the main text, which suggests a thickness of 6.29 nm there is an overestimation of the thickness by a factor of more than 2. This could be a result of the PPC residues as can be seen in both topography scans of Fig. \ref{suppfig:AFM}a and b and an overestimation of the van der Waals gap due to water and/or dirt as described by Shearer \textit{et al. } \cite{shearer2016accurate}. In Fig. \ref{suppfig:AFM}c\&d, AFM scans of the same device after annealing are shown. It can be seen in the inset of Fig. \ref{suppfig:AFM}c that the line scan across the edge of the SRO flake is now 10.9 nm which is still an overestimate compared to the 6.29 nm we expect from the XRD data but is reduced from the data taken before annealing. Annealing can remove a large portion of the PPC residues. Also, a reduction in the amount of buckling is seen in Fig. \ref{suppfig:AFM}d. Before annealing, the SRO membrane shows larger buckling amplitudes (Fig. \ref{suppfig:AFM}b), likely arising from the PPC stamping process. However, after annealing the buckling height is largely reduced. 

\clearpage
\section{Exploded drum}
\begin{figure}
\centering
\includegraphics[width=\columnwidth]{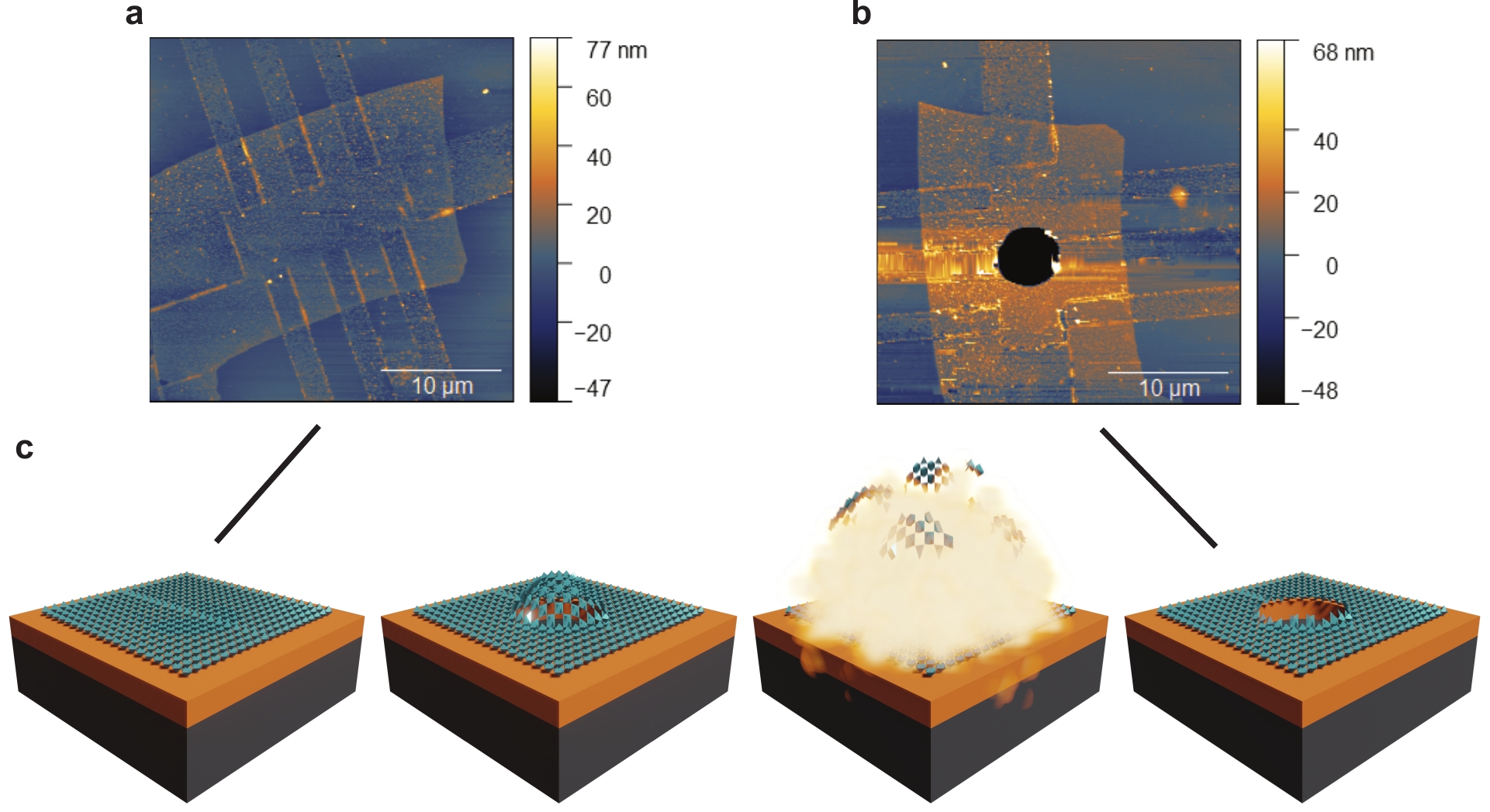}
\caption{Atomic force microscopy (AFM) images of an annealed SRO flake \textbf{a} before loading into a vacuum chamber and \textbf{b} after loading into a vacuum chamber. \textbf{c} Illustration of the explosion process in annealed SRO drum. In ambient conditions, SRO flake is in the flat configuration. When the sample chamber is abruptly pumped, the membrane bulges up due to the pressure difference between the cavity and the sample chamber. At base pressure of $\sim$10$^{-3}$ bar, the membrane bursts leaving a SRO flake with a hole in the center.}
\label{suppfig:explosion}
\end{figure}

Figure \ref{suppfig:explosion} shows atomic force microscopy (AFM) images of a SRO flake on cavities of \ch{SiO2}/Si with pre-patterned electrodes. Shown device is annealed at 300$^\circ$C for 15 minutes. After annealing the device is bonded and loaded into a Oxford He flow cryostat. Just prior to loading, the membrane is in a flat configuration as shown in Fig. \ref{suppfig:explosion}a. In the process of loading, the membrane bulges up due to the pressure difference between the cavity, which is filled with $\sim$1 bar air and the cryostat which is pumped to $\sim$1 mbar before letting the He to flow. After the attempt to measure the electron transport in the cryogenic temperatures we removed the sample and observed the drum using AFM. As shown in Fig. \ref{suppfig:explosion}b, the drum had been torn off from the device. We hypothesize that this has occurred in a violent fashion as illustrated in Fig. \ref{suppfig:explosion}c. During the pumping phase, the membrane may have ``exploded'' from the device due to a sudden change in the pressure. Having sealed, the pressure difference between the cavity and the sample space in the cryostat will have had $\Delta P$ $\sim$ 1 bar. 

\clearpage
\section{Ultra-fast acoustics}

\subsection*{Method}
Acoustic pulses in the GHz range are generated and detected using a picosecond ultrasonics technique \cite{thomsen1984coherent}. By locally heating the sample with a 100 fs pump pulse, an acoustic pulse is generated. Using a probe pulse of the same duration, the acoustic reflections arriving back at the surface of the sample are measured. In this pump-probe experiment, the acoustic echoes are measured with the probe repetition rate at a slight offset from the 100 MHz pump, regulated by an asynchronous optical sampling (ASOPS) system from Menlo Systems. The pump and probe lasers are femtosecond erbium lasers (1560 nm), doubled in frequency (780 nm) for the probe. The average output power of the pump laser is $\sim$100 mW, and the one of the probe laser is set between 2 and 5 mW. Since the pump and probe repetition rates have a slight offset between them, at every next pulse the time of arrival between the pump and probe is slightly delayed in time. This way, the probe scans through the entire time window between 2 pump pulses (10 ns).

\begin{figure}
\centering
\includegraphics[width=0.8\columnwidth]{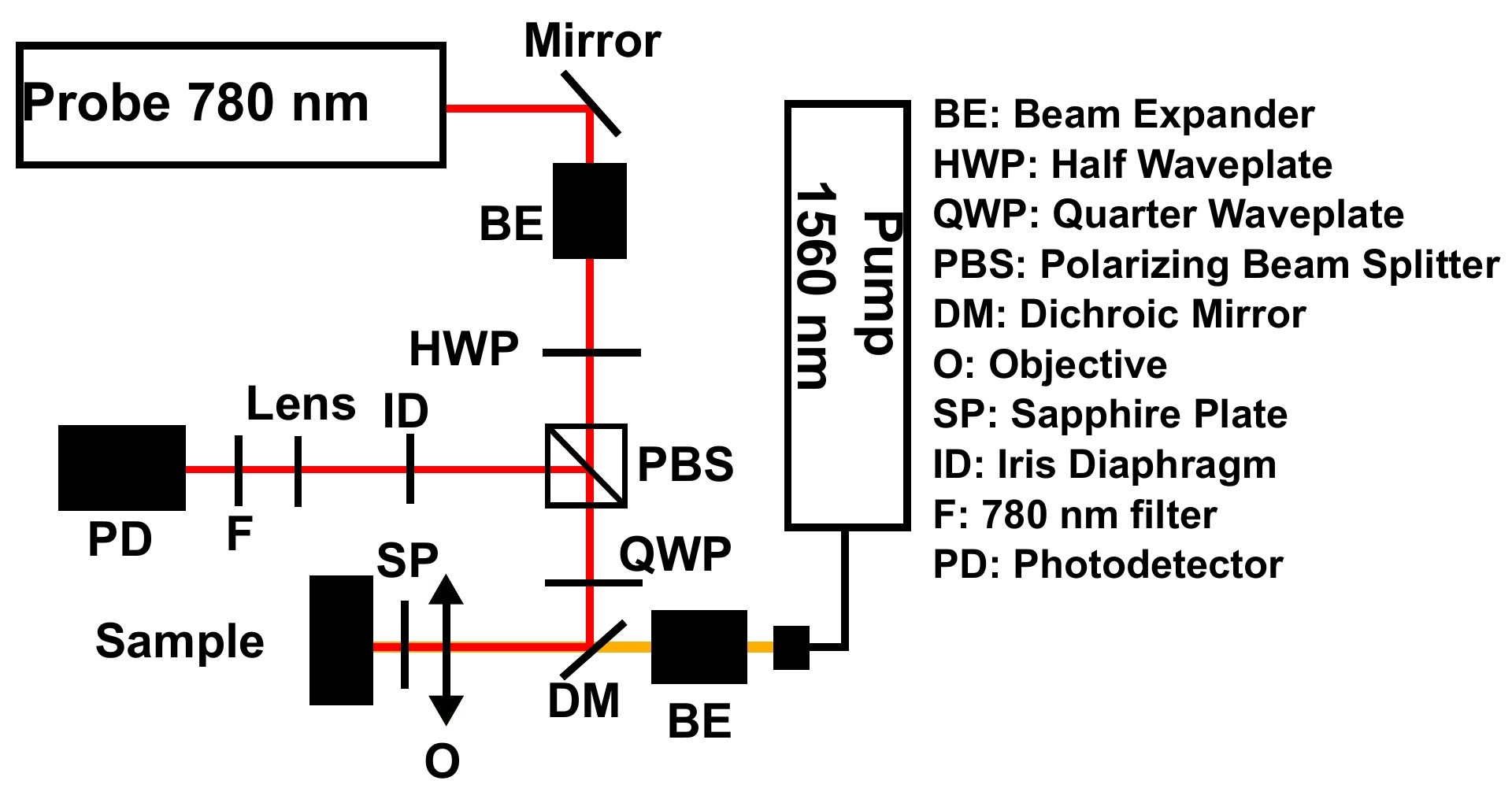}
\caption{Schematic illustration of the picosecond ultrasonics setup.}
\label{suppfig:uf_setup}
\end{figure}

In the setup (Fig. \ref{suppfig:uf_setup}), the probe beam is expanded to an appropriate size and passed through a half waveplate to achieve the correct polarization to be transmitted by the polarizing beam splitter. A quarter waveplate then shifts the initial linear polarization to an elliptical polarization. After passing through a dichroic mirror, where the pump path joins the probe path, the beams are focused on the sample through a sapphire plate, used for ultrasonics detection by means of conoscopic interferometry \cite{liu2018common}. On the sample, the pump beam is absorbed to generate the acoustic waves. A part of the probe beam containing the acoustic signal is reflected. Then the quarter waveplate ensures that the reflected path is reflected towards the photodetector by the polarizing beam splitter. In this path, an iris diaphragm cuts out a part of the light, which is necessary for the conoscopic interferometry. A 250 MHz bandwidth, silicon based, amplified photodetector is used in this setup, allowing to detect each individual pulse.

To acquire the data we use a 600 MHz lock-in amplifier from Zurich Instruments with the additional Boxcar option. This option allows to combine a Periodic Waveform Analyser (PWA) and a Boxcar. Using the boxcar, the energy contained inside each detected probe pulses is extracted. From this, the PWA reconstructs the ASOPS signal by placing the pulses in the correct order. The sampled data are linked to the phase of an oscillator at the probe repetition rate. By plotting the sample amplitude against this phase, the laser pulses can be reconstructed (PWA). Using a gate window this pulse can now be integrated and the amplitude can be derived (Boxcar). A second oscillator at the offset frequency between the pump and probe lasers, is also linked to the measurement. Using the phase of this second oscillator, the samples are divided in bins that are associated to corresponding time stamp within the measurement at which the sample was taken. By calculating the amplitude in each of this bins using the boxcar method and then concatenating them, the envelope signal of all the pulses is reconstructed.

\subsection*{Raw data}
\begin{figure}
\centering
\includegraphics[width=\columnwidth]{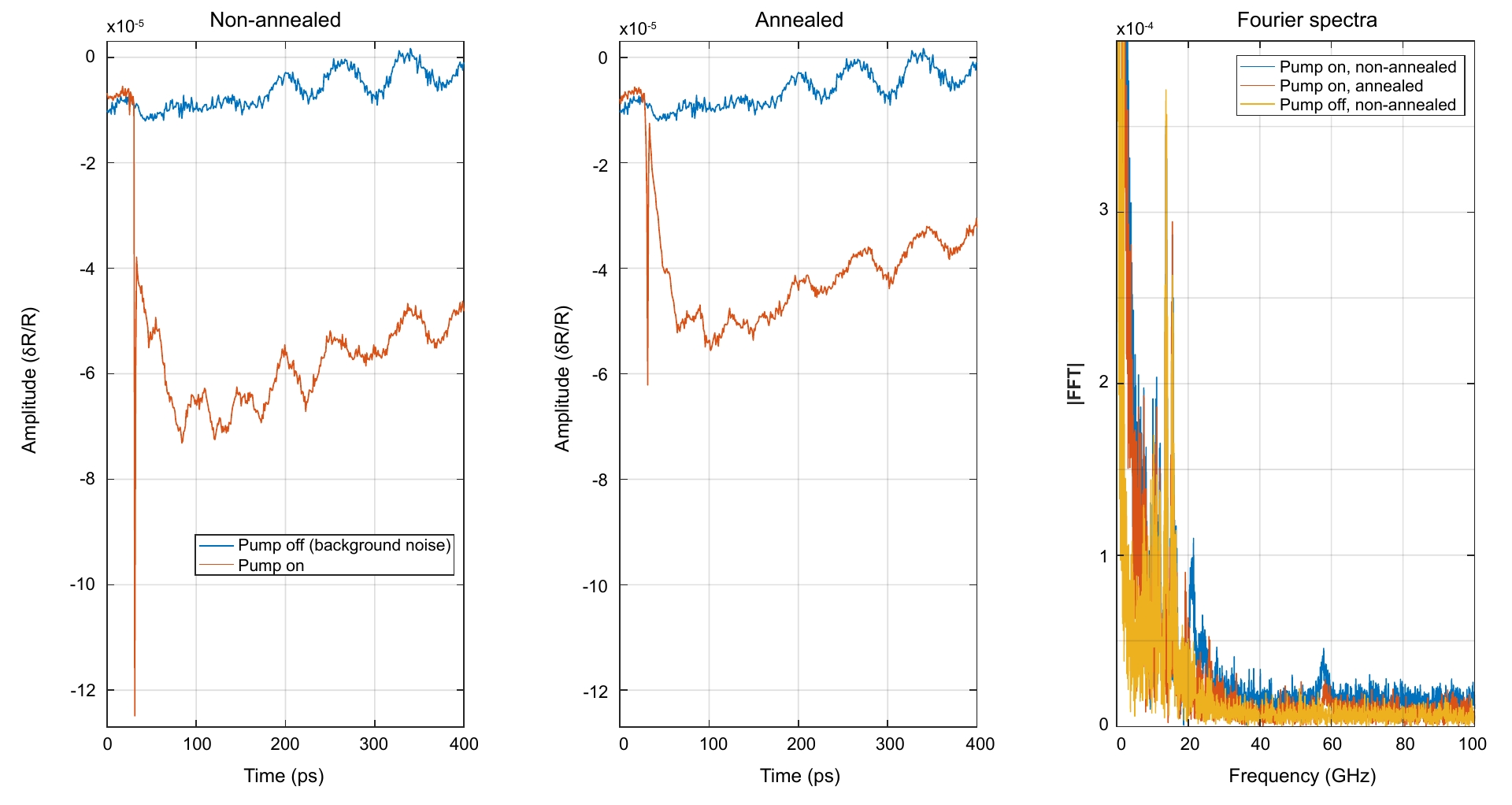}
\caption{Raw signal (zoom around the excitation). \textbf{Left:} Non-annealed sample. \textbf{Middle:} Annealed sample. \textbf{Right:} Fourier transformed spectra.}
\label{suppfig:uf_raw}
\end{figure}

In the raw signal (Fig. \ref{suppfig:uf_raw}), increases in the sample temperature caused by the pump laser pulses can be seen in the left and middle graph after a few ps. The effect is much higher in the non-annealed sample. This could be explained by the adhesion to the substrate. In the annealed sample, the improved adhesion facilitates a more efficient heat diffusion into the substrate, while in the non-annealed sample, it stays more confined inside the Au/Cr/\ch{SrTiO3} assembly.

After $\sim$200 ps, a parasitic signal can be seen (also present without the pump excitation). This parasitic signal and the temperature increase are the origin of the high level component under $\sim$20 GHz in the spectra (Fig. \ref{suppfig:uf_raw} right). The part of the signal between the temperature increase and the parasitic signal ($\sim$50 ps – 200 ps) correspond to the acoustic signal, which is partially masked by the parasitic signal, as it can be seen in the non-annealed sample. The components of the spectra around $\sim$20 GHz and $\sim$57 GHz correspond to the frequency of these acoustic standing waves which are absent from the spectrum of the signal without excitation (Fig. \ref{suppfig:uf_raw} right, plotted in yellow) and are weaker in the annealed sample (Fig. \ref{suppfig:uf_raw} right, plotted in orange).

\subsection*{Filtered data}
\begin{figure}
\centering
\includegraphics[width=\columnwidth]{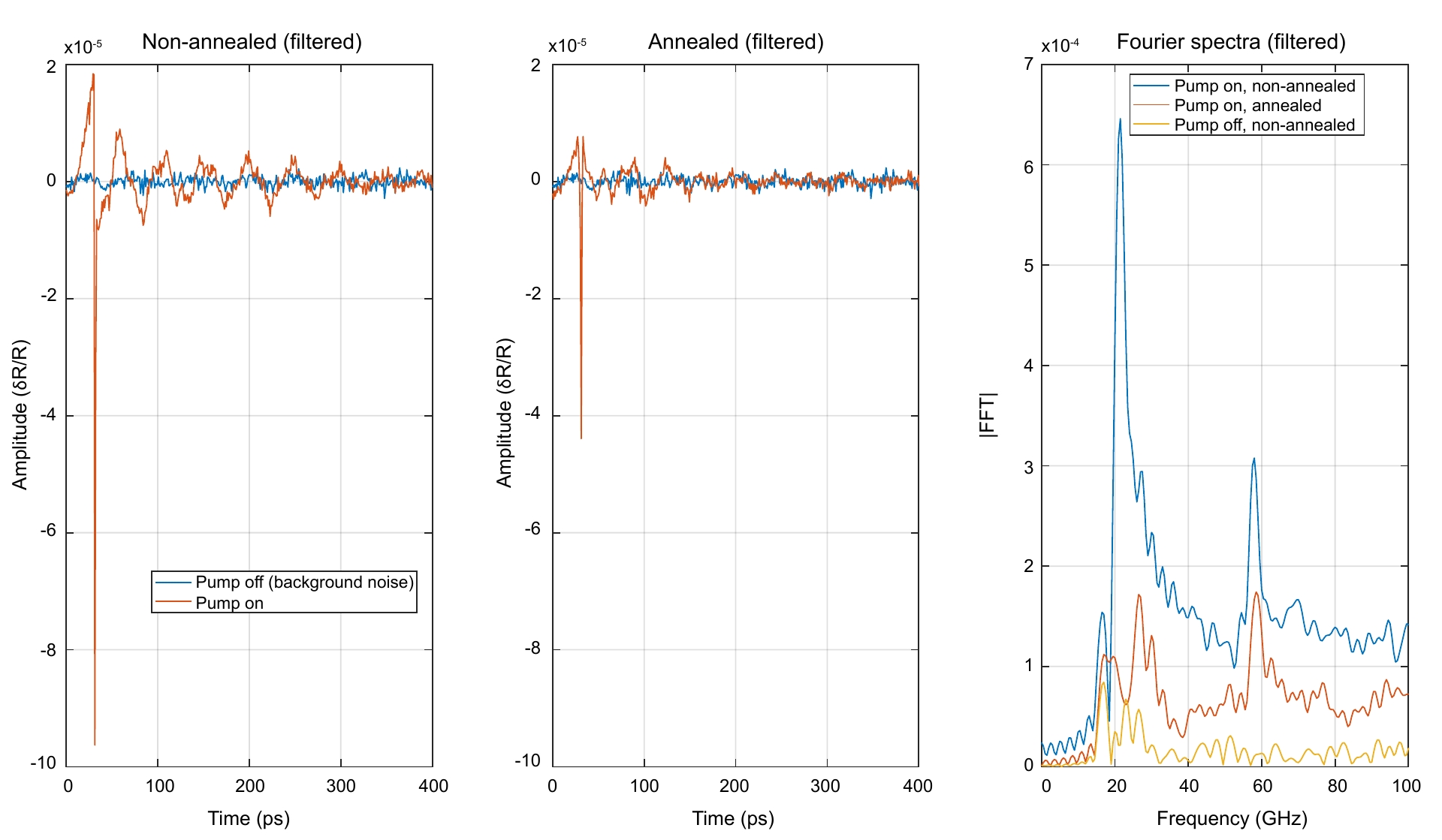}
\caption{High-pass filtered signal. \textbf{Left:} Non-annealed sample. \textbf{Middle:} Annealed sample. \textbf{Right:} Fourier transformed spectra.}
\label{suppfig:uf_filtered}
\end{figure}

The signals are then high-pass filtered using a 18 GHz cut-off frequency to remove the parasitic background signals and the low frequency components of the temperature increase (Fig. \ref{suppfig:uf_filtered}). The acoustic waves become much more visible, as well their associated components in the Fourier spectra.

\subsection*{Fitting a sinusoidal envelope}
\begin{figure}
\centering
\includegraphics[width=\columnwidth]{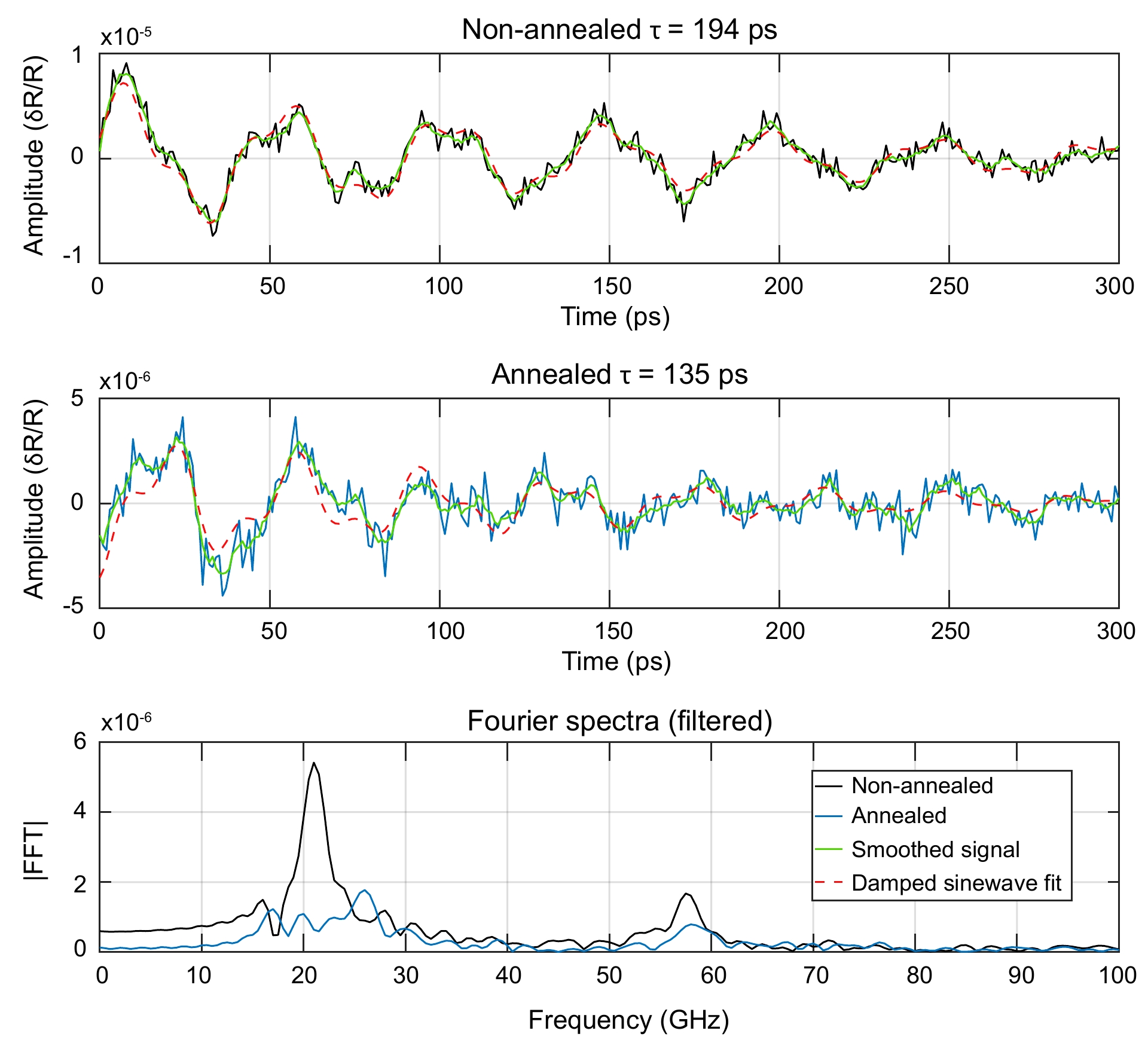}
\caption{Fit of the signals. \textbf{Top:} Non-annealed sample. \textbf{Middle:} Annealed sample. \textbf{Bottom:} Fourier transformed spectra. }
\label{suppfig:uf_fit}
\end{figure}

Only the part of the signal corresponding to the acoustic waves is then considered. This one is firstly smoothed using a moving average filter on 5 samples and is then fitted using a minimisation algorithm to find the parameters of a damped sine as the following (Eq. \ref{suppeq:sin}):

\begin{equation}
\text{fit}(t)=(A_1\sin(2\pi f_1t+\phi_1)+A_2\sin(2\phi_2t+\phi_2))e^{-t/\tau_{ac}},
\label{suppeq:sin}
\end{equation}
where $f_1$ and $f_2$ are the frequencies of the first and second components in the signal ($\sim$20 GHz and $\sim$57 GHz respectively), both determined using the spectrum of Fig. \ref{suppfig:uf_filtered}, $A_1$ and $A_2$ are the amplitude of both components and $\phi_1$ and $\phi_2$ their phases. Finally, $\tau_{ac}$ is the time constant that characterizes the loss in acoustic amplitude due to consecutive reflections at the boundary, by transmitting this acoustic energy to the substrate. With good adhesion between the \ch{SrTiO3} and \ch{SiO2}, a relatively large part of the acoustic energy is transmitted to the substrate, inducing a low value of $\tau_{ac}$. For a bad adhesion, where most of the acoustic energy is kept inside the Au/Cr/\ch{SrTiO3} assembly, the time constant is larger. Theoretically, two different time constants should be considered, one for each frequency ($\tau_1$ and $\tau_2$). However, the differences between $\tau_1$ and $\tau_2$ is quite weak ($\sim$10-20 ps) at the frequencies considered here and are below the error threshold on the estimation of the value of $\tau_{ac}$. Furthermore, adding one more parameter for the fit decreases the reliability of the results. Therefore, the approximation $\tau_1$ $\approx$ $\tau_2$ = $\tau_{ac}$ is considered here. The attenuation of the acoustic waves during the propagation inside the structure due to other effects (viscosity, scattering, etc.) is neglected here. The result is presented on Fig. \ref{suppfig:uf_fit} for both non-annealed and annealed case with their associated spectra. The black and blue curves correspond to the filtered raw data of non-annealed and annealed samples respectively, the green to the smoothed signal and the dashed red lines to the fit using Eq. \ref{suppeq:sin}.

\subsection*{Calculations of the results}

The theoretical resonance frequencies are calculated from the expressions of standing waves in a medium, as given by Greener \textit{et al.} \cite{greener2019high}:
\begin{equation}
f_{nU}=\frac{nc_{Leq}}{2h_{eq}},
\label{suppeq:unbound}
\end{equation}
for an unbounded medium (total debonding case, i.e. free surface at each boundary) and
\begin{equation}
f_{nB}=\frac{(2n-1)c_{Leq}}{4h_{eq}},
\label{suppeq:bound}
\end{equation}
for a bounded medium (perfect adhesion case, i.e. loaded surface with continuity of stresses and displacement at the boundary with the substrate and free surface boundary condition at the other boundary of the structure). In the Eqs. \ref{suppeq:unbound} and \ref{suppeq:bound}, $n$ is the order of the harmonic, $c_L{eq}$ = $\frac{c_{L1} c_{L2} h_{eq}}{c_{L1} h_2 + c_{L2} h_1}$ is the equivalent longitudinal wave velocity in the Au/Cr/\ch{SrTiO3} assembly with $c_{L1}$ and $c_{L2}$ respectively the longitudinal velocity in the Au and \ch{SrTiO3} layer. $h_{eq}$ = $h_1$ + $h_2$ is the total thickness of the assembly with $h_1$ = 30 nm the thickness of the Au layer and $h_2$ = 80 nm the one of the \ch{SrTiO3} layer. The chromium layer is here neglected due to its small thickness (3 nm) with respect to the total thickness of the structure and with respect to the acoustic wavelength (of the order of 100 nm).

The reflection coefficient is then deduce from the following formula, also used by Greener \textit{et al.} \cite{greener2019high}:
\begin{equation}
|A_p|=|A_0||R_{ac}|^p=|A_0|\left(e^{-\frac{2h_{eq}}{c_{Leq}\tau_{ac}}}\right)^p=|A_0|\left(e^{-\frac{1}{\tau_{ac} f_{1U}}}\right)^p,
\label{suppeq:greener2}
\end{equation}
where $|A_p|$ is the absolute value of the amplitude after p reflections of the acoustic wave inside the structure, $|A_0|$ the initial amplitude of the wave, $|R_{ac}|$ the absolute value of the reflection coefficient in amplitude. From the reflection coefficient, it is possible to deduce the longitudinal interfacial stiffness $K_L$, characterizing the adhesion between both materials at the interface \cite{grossmann2017characterization}:

\begin{equation}
|R_{ac}| = \left|\frac{Z_2-Z_s+\frac{j\omega Z_2 Z_s}{K_L}}{Z_2+Z_s+\frac{j\omega Z_2 Z_s}{K_L}}\right|,
\label{suppeq:Rac}
\end{equation}
where $Z_m$ = $c_{Lm}\rho_{m}$ the acoustic impedance of the medium $m$ with $\rho_m$ its density. Here, $m$ = 2 for the \ch{SrTiO3} and $m$ = $s$ for the substrate (\ch{SiO2}). For all these calculations, the values of $c_{L1}$, $c_{L2}$, $c_{Ls}$, $\rho_{2}$, $\rho_{s}$ have been taken from literature \cite{uozumi1972sound, muta2005thermoelectric, brick2017picosecond} and are given in Table \ref{supptable:values}

\begin{table}
\footnotesize
\centering
\begin{tabular}{|c | c| c| c| c|}
\hline
\textbf{Parameter} & \textbf{Au} & \textbf{\ch{SrTiO3}} & \textbf{Equivalent (Au/\ch{SrTiO3})}& \textbf{\ch{SiO2}}\\ \hline
\textbf{Longitudinal velocity $c_L$ (m/s)} & 2200 &7900 & 4630 & 5800\\ \hline
\textbf{Density $\rho$(kg/m$^3$)}&-&5110&-&2650\\ \hline
\end{tabular}
\label{supptable:values}
\caption{Values of the longitudinal wave velocities and densities of the different materials.}
\end{table}

\begin{figure}
\centering
\includegraphics[width=\columnwidth]{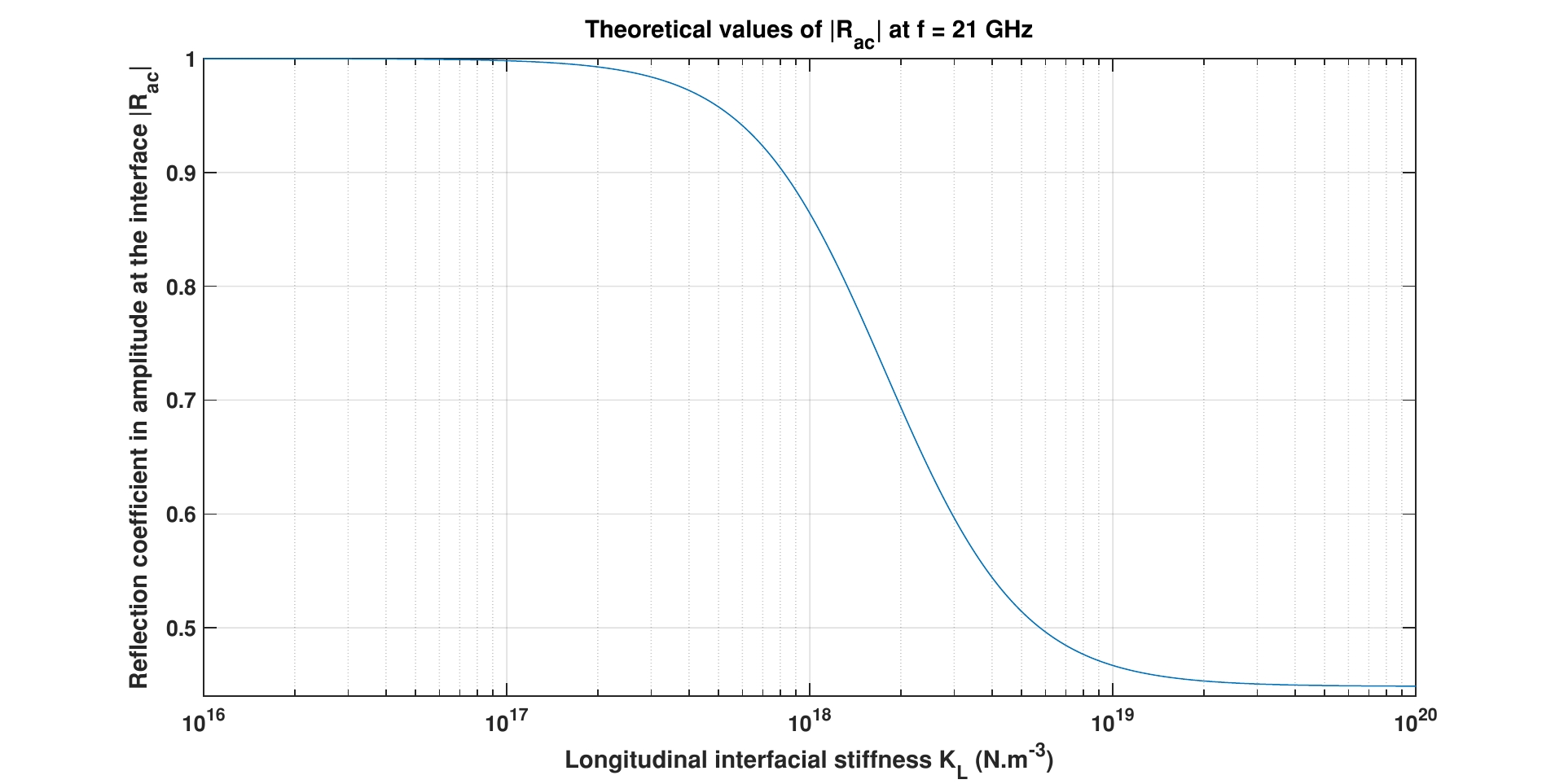}
\caption{Variations of the reflection coefficient with respect to the longitudinal interfacial stiffness.}
\label{suppfig:uf_R-KL}
\end{figure}

Figure \ref{suppfig:uf_R-KL} shows the variations of the reflection coefficient in amplitude by varying the interfacial stiffness at the frequency $f_{1U}$ = 21 GHz.

The picosecond ultrasonics measurements were performed on 5 different annealed flakes and 4 different non-annealed flakes. Table \ref{supptable:results} gives the values of the frequencies and time constants for each measurements.

\begin{table}
\footnotesize
\centering
\begin{tabular}{l|ccc|ccc}
                   & \multicolumn{2}{l}{\textbf{Annealed flakes}} &       & \multicolumn{3}{l}{\textbf{Non-annealed flakes}} \\ \hline
                   & \textbf{$f_{1u}$ (GHz) }&\textbf{$f_{3B}$ (GHz) } & \textbf{$\tau_{ac}$ (ps)} &  \textbf{$f_{1u}$ (GHz) }  & \textbf{$f_{3B}$ (GHz)  } & \textbf{$\tau_{ac}$ (ps)}            \\ \hline
\textbf{1 }                 & 26               & 58               & 135.3 & 21          & 57.5        & 194.1       \\ \hline
\textbf{2 }                 & 22.5             & 56               & 127.2 & 22.5        & 53          & 225.4       \\ \hline
\textbf{3 }                 & 22.5             & 58               & 97.6  & 23          & 57          & 287.4       \\ \hline
\textbf{4 }                 & 25               & 57               & 109.9 & 21          & 53.5        & 172.7       \\ \hline
\textbf{5 }                & 27.5             & 60               & 98.2  & -           & -           & -           \\ \hline \hline
\textbf{Avg}           & 24.7             & 57.8             & 113.6 & 21.9        & 55.3        & 219.9       \\ \hline
\textbf{Std} & 2.2              & 1.5              & 17.1  & 1           & 2.3         & 50         \\ \hline
\end{tabular}
\label{supptable:results}
\caption{Values of the frequencies and time constants for each measurements on different flakes.}
\end{table}

\subsection*{COMSOL simulations}

\begin{figure}
\centering
\includegraphics[width=\columnwidth]{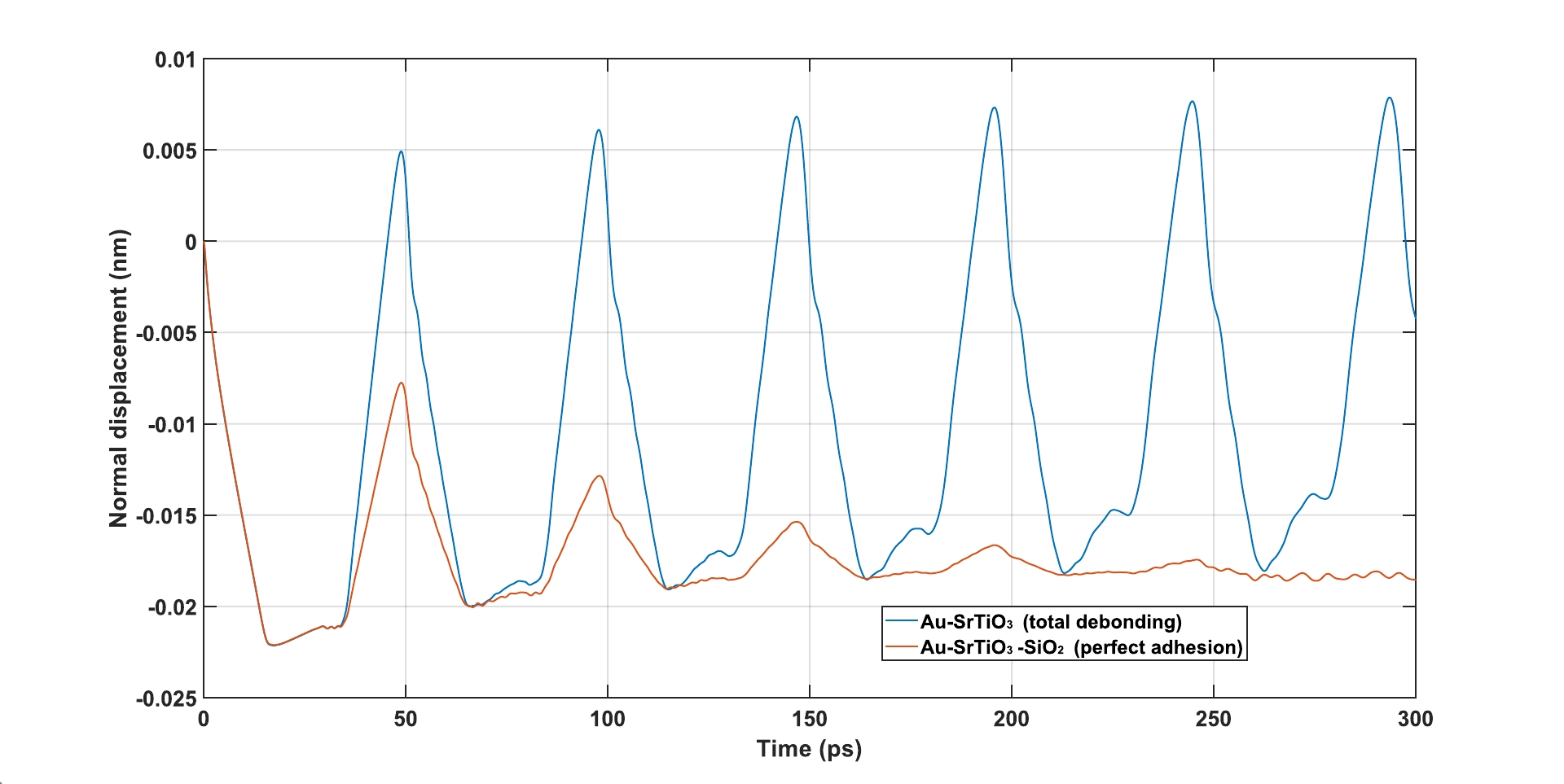}
\caption{Comsol simulation: comparison between the total debonding (blue) and perfect adhesion cases (orange).}
\label{suppfig:uf_comsol}
\end{figure}

The Finite Element Method simulations were performed using the heat transfer in solid and solid mechanics modules of the software COMSOL Multiphysics. The two modules were coupled together using the multiphysics thermal expansion module. The laser excitation was modeled as a heat source with similar characteristics than the pump laser pulses. The acoustic waves were detected using a point probe measuring the normal displacement at the surface of the 2D sample.

To simulate the total debonding, only an Au/\ch{SrTiO3} sample was considered using free boundary condition at each end. To simulate a perfect adhesion case, an \ch{Au/SrTiO3/SiO2} sample was considered with a ``perfect'' interface (corresponding to a case where $K_L$ $\rightarrow$ $\infty$ in Eq. \ref{suppeq:Rac} ).

A triangular meshing was used with a maximum element size of 2.5 nm in Au and 4 nm in \ch{SrTiO3} to have minimum $\sim$15 elements per wavelength at 60 GHz in each materials. A coarser meshing (10 nm) was used in the \ch{SiO2} substrate since the propagation of the waves in this material is not really of interest here. Its thickness has been chosen large enough to not have any backward reflection coming from the bottom of the substrate. A time dependent study is then processed with a time step of 0.2 ps to get more than 60 samples per period of the acoustic waves at 60 GHz.

The results of the simulations are shown in Fig. \ref{suppfig:uf_comsol}. For the case with the substrate (corresponding to a perfect adhesion, orange curve on the Fig. \ref{suppfig:uf_comsol}), the reflection coefficient in amplitude $|R_{ac}|$ of 0.45 is found: this case corresponds to the limit $K_L$ $\rightarrow$ $\infty$, as shown in the Fig. \ref{suppfig:uf_R-KL}. The case without the substrate (corresponding to a total debonding, blue curve in the Fig. \ref{suppfig:uf_comsol}), the reflection coefficient is equal to 1: this case corresponds to the limit $K_L$ $\rightarrow$ 0, as shown in Fig. \ref{suppfig:uf_R-KL}.

\clearpage

\bibliography{supp_selfsealing}
\bibliographystyle{ieeetr}